\documentclass[aps,prx,twocolumn,unsortedaddress,floatfix,nofootinbib,superscriptaddress]{revtex4-2}
\usepackage{amsmath}
\usepackage{amssymb}
\usepackage{bbm}
\usepackage{mathtools}
\usepackage{physics}
\usepackage{graphicx}
\usepackage{dcolumn}
\usepackage{bm}
\usepackage{braket}
\usepackage{qcircuit}
\usepackage{svg}
\usepackage[colorlinks]{hyperref}
\usepackage{siunitx}

\let\oldaddcontentsline\addcontentsline% Store \addcontentsline
\renewcommand{\addcontentsline}[3]{}% Make \addcontentsline a no-op

\newcommand{\bseq}{\begin{subequations}}
\newcommand{\eseq}{\end{subequations}}

\definecolor{linkcolor}{RGB}{0,83,166}
\hypersetup{
  colorlinks = true,
  allcolors = {linkcolor}
}

\usepackage{xcolor}
\usepackage{soul}
\usepackage{ulem}

\newcommand{\dwave}{\affiliation{D-Wave Quantum, 3033 Beta Ave., Burnaby, BC V5G 4M9, Canada}}
\newcommand{\usc}{\affiliation{Department of Physics and Astronomy, University of Southern California, Los Angeles, CA 90089, USA}}
\newcommand{\psis}{\affiliation{Paul Scherrer Institute, 5232 Villigen PSI, Switzerland}}
\newcommand{\cqist}{\affiliation{Center for Quantum Information Science \& Technology, University of Southern California, Los Angeles, CA 90089, USA}}
\newcommand{\sfu}{\affiliation{
Department of Physics, Simon Fraser University, Burnaby, BC V5A 1S6, Canada}}
\newcommand{\nmc}{\affiliation{New Mexico Consortium, Los Alamos, NM 87544, USA}}
\begin{document}

\preprint{APS/123-QED}

\newcommand{\mytitle}{Analog-Digital Quantum Computing with Quantum Annealing Processors}
\title{\mytitle}

\author{Rahul Deshpande}
 \email{rdeshpande@dwavesys.com}
\dwave

\author{Majid Kheirkhah}
\dwave
\sfu

\author{Chris Rich}
\dwave

\author{Richard Harris}
\dwave

\author{Jack Raymond}
\dwave

\author{Emile Hoskinson}
\dwave

\author{Pratik Sathe}
\dwave
\nmc

\author{Andrew J. Berkley}
\dwave

\author{Stefan Paul}
\dwave

\author{Brian Barch}
\usc
\cqist

\author{Daniel A. Lidar}
\usc
\cqist
\affiliation{Departments of Chemistry and Electrical \& Computer Engineering, University of Southern California, Los Angeles, CA 90089, USA}
\affiliation{Quantum Elements, Inc., 2829 Townsgate Road, Westlake Village, CA 91361, USA}

\author{Markus Müller}
\psis

\author{Gabriel Aeppli}
\psis
\affiliation{
Department of Physics and Quantum Center, Eidgenossische Technische Hochschule Zürich, 8093 Zürich, Switzerland}
\affiliation{
Institut de Physique, Ecole Polytechnique Fédérale de Lausanne, 1015 Lausanne, Switzerland}

\author{Andrew D.~King}
\dwave

\author{Mohammad H. Amin}
\dwave
\sfu

\date{\today}

\begin{abstract}
Quantum annealing processors typically control qubits in unison, attenuating quantum fluctuations uniformly until the applied system Hamiltonian is diagonal in the computational basis.  This simplifies control requirements, allowing annealing QPUs to scale to much larger sizes than gate-based systems, but constraining the class of available operations. Here we expand the set of available operations by demonstrating analog-digital quantum computing, including the measurement of spatio-temporal correlation functions, in a large-scale superconducting quantum annealing processor. This involves evolution under a fixed many-body Hamiltonian that, in the weak-coupling regime, is well described by an effective XY Hamiltonian, together with arbitrary-basis initialization and measurement via auxiliary qubits. Operationally, this is equivalent to implementing single-qubit gates before and after an analog quantum evolution. We illustrate this capability with single-qubit and two-qubit coherent oscillations with varying initialization and measurement bases. We then demonstrate fermionic dispersion in a periodic spin chain and Anderson localization, including the predicted quadratic scaling of the imbalance with disorder, both in excellent agreement with theory. Our work shows that commercial quantum annealers can be operated as analog-digital quantum computers with sufficient coherence to simulate dynamics of large-scale quantum many-body systems, opening the door to a wide range of new applications.
\end{abstract}

\maketitle

Simulating the dynamics of quantum systems, which are inherently analog, is one of the most promising applications of today's quantum computers~\cite{barends_digitized_2016, zhang_observation_2017, scholl_quantum_2021, ebadi_quantum_2021, king_quantum_2023, kim_evidence_2023, shaw_benchmarking_2024, miessen_benchmarking_2024, king_beyondclassical_2025, AndersenThermalization2025a, manovitz_quantum_2025, haghshenas_digital_2025, lunkin_evidence_2026}.  Among quantum computing modalities, superconducting quantum annealing (QA) processors~\cite{harris_experimental_2010,johnson_quantum_2011} have the advantage of large scale, complex connectivity and readily tunable analog quantum evolution. These capabilities are made possible via highly multiplexed control, which comes at the cost of flexibility compared to today's small-scale gate-based systems.  Recent studies have demonstrated the utility of analog Hamiltonian evolution within gate-based systems~\cite{shaw_benchmarking_2024, AndersenThermalization2025a, lunkin_evidence_2026}; this paradigm avoids the overhead and inaccuracy associated with Trotterizing analog evolution into a set of discrete gates~\cite{barends_digitized_2016, kim_evidence_2023, miessen_benchmarking_2024}.  Here we address the converse question: Can the flexibility of gate-based state preparation and measurement be realized within an annealing-based system?

To answer this question, we implement an analog evolution under a programmable time-independent, many-body Hamiltonian bracketed by two additional layers comprising individually-programmable, single-qubit rotations. Equivalently, we use a layer of single-qubit gates to initialize the system in a chosen state, greatly expanding the state space accessible on quantum annealing processors and enabling the study of coherent excited-state evolution.  The second layer of single-qubit gates can rotate the measurement basis to any arbitrary basis, which in principle is sufficient for full tomographic reconstruction of the final quantum state.

This implementation is a form of analog-digital quantum computing (ADQC), combining the large-scale analog simulation of quantum dynamics with rapid gate-based (digital) initialization and readout steps. Extended versions of this protocol have been proposed as pathways to universal quantum computing
~\cite{Parra-RodriguezDigitalanalog2020,imoto_2025} and efficient solution of classically intractable problems~\cite{lidar_digitalanalogdigital_2025}.  Experimental realizations have been demonstrated in trapped ions~\cite{KatzHybrid2025}, superconducting circuits~\cite{KumarDigitalanalog2025,AndersenThermalization2025a}, and neutral atoms \cite{bluvstein_quantum_2022,geim_engineering_2026}. Here, we achieve this capability on a thousand-qubit-scale quantum annealing processor and demonstrate many-body quantum dynamics with coherent behavior up to normalized time $|\mathcal{J}|\,t/\hbar \gtrsim 100$, where $\mathcal{J}$ is the coupling energy. This implies that quantum information can propagate across more than a hundred qubits in any spatial direction before coherence is lost~\cite{SM}.

\begin{figure*}
\includegraphics[scale=1]{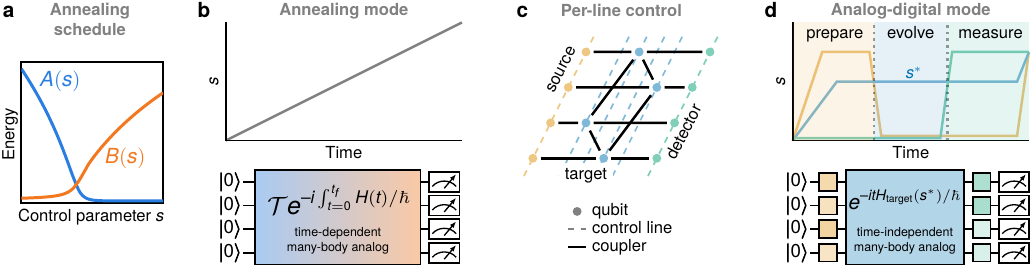}
\caption{{\bf Analog-digital protocol in a quantum annealer.} 
{\bf a}, The time-dependent quantum annealing (QA) Hamiltonian $H(t)$~\eqref{eq:H_anneal} is governed by transverse and Ising energy scales $A(s)$ and $B(s)$ and a control parameter $s(t)$. 
{\bf b}, In standard QA, $s(t)$ increases linearly from $0$ to $1$, interpolating $H(t)$ from a transverse-field-dominated Hamiltonian to one dominated by the problem Hamiltonian. 
{\bf c}, Anneal schedules $s^\alpha(t)$ are controlled independently on several annealing lines $\alpha$, not necessarily in unison. Our analog-digital setup reserves one or more lines for source and detector qubits, which respectively prepare and measure the states of the remaining ``target'' qubits. 
{\bf d}, Free evolution of the target qubits under the action of a time-independent many-body analog Hamiltonian $H_\text{target}(s^*)$ is initiated by a quench of the source qubits to $s=0$ and terminated by a quench of the detector qubits to $s=1$; operationally, these processes implement individually programmable single-qubit gates before and after the evolution under $H_\text{target}(s^*)$, enabling very fast (on the timescale
of target qubit interactions) initialization and readout.}\label{fig:1}
\end{figure*}
\section*{\label{sec:theory_sim}Implementation on Quantum Annealing processors}

Quantum annealing  processors implement programmable Hamiltonians with static single-qubit and two-qubit terms and global, time-dependent prefactors. The static parameters can be set using on-chip programmable magnetic memories controlled by a multiplexed addressing scheme, enabling the specification of $n$ static parameters using only $\mathcal{O}(n^{1/3})$ low-frequency bias lines~\cite{Bunyk2014}. The global fields can be applied using low-frequency lines shared by a large number of qubits. This has enabled QA processors to scale to thousands of qubits using only hundreds of control lines.

In the processor used in this work, there are six low-frequency ($\SI{30}{MHz}$ low-pass-filtered) annealing lines, each controlling the anneal parameter $s$ for an independent set of qubits. We highlight this independent control by introducing the per-qubit time-dependent anneal parameters $s_i(t) = s^{\alpha(i)}(t) + s_{i0}$, where $\alpha(i) \in \{1, \dots, 6\}$ denotes the global line controlling the qubit and $\{s_{i0}\}$ are programmable ``anneal offsets''
 \cite{Lanting_2017_nonuniform_driver}.
The Hamiltonian of the system can be written as
\begin{align}
\label{eq:H_anneal}
H(t)=& \,
-\frac{1}{2} \sum_i A\big( s_i(t) \big) \tau^x_i + \frac{1}{2} H_P(t), 
\end{align}
where (dropping the explicit time-dependence of $s_i$ for simplicity)
\begin{align}\label{eq:HP_anneal}
H_P (t) =&
\sum_{\langle i,j\rangle} \sqrt{B(s_i)B(s_j)} J_{ij}\tau^z_i\tau^z_j
+ \sum_i B(s_i) h_i\tau^z_i.
\end{align}
Here, $\tau^{x,z}_i$ are Pauli matrices acting on qubit $i$; and the longitudinal fields $h_i$ and exchange interactions $J_{ij}$ are individually programmable (in this work, we set ${h_i=0}$). The functional forms of the energy scales $A(s)$ and $B(s)$
are shown in Fig.~\ref{fig:1}(a). In addition to the programmable parameters 
$h_i$ and $J_{ij}$, the system allows for per-qubit programmable static flux biases. These act analogously to the local fields $h_i$ but are not subject to time-dependent prefactors; equivalently, they contribute additional longitudinal-field terms not written explicitly in Eqs.~\eqref{eq:H_anneal} and \eqref{eq:HP_anneal}. As we shall see, static flux biases play a crucial role in enabling single-qubit gate operations within the ADQC framework.

In standard QA protocols, all lines are synchronized so that $s_i(t) = s(t)$. In a typical forward-anneal protocol \cite{Boixo2014a, king_quantum_2023, king_beyondclassical_2025}, the anneal parameter evolves linearly as 
$s(t)=t/t_f$ and the time-dependent Hamiltonian 
$H(t)$ is governed by the processor-specific anneal schedules $(A(s),B(s))$ (Fig.~\ref{fig:1}a--b). Depending on the annealing time $t_f$, the system ends up in a low-energy state of the problem Hamiltonian $H_P(t_f)$, thereby enabling quantum optimization~\cite{farhi_quantum_2001,albash_adiabatic_2018}. More general forms of $s(t)$ are available, for example pause-and-quench protocols with $s(0)=0$ and $s(t_f)=1$, or reverse-anneal protocols with $s(0)=s(t_f)=1$~\cite{harris_phase_2018, king_observation_2018}.

Here we further generalize this control by introducing ``multicolor annealing'', where we allow a distinct control parameter $s^\alpha(t)$ on each line.  This enables us to group the qubits into three sets: ``source'', ``target'', and ``detector'' qubits, whose lines are controlled with independent anneal parameters $s^\text{source}(t)$, $s^\text{target}(t)$, and $s^\text{detector}(t)$, respectively (Fig.~\ref{fig:1}c). We allocate four lines to the target qubits, which encode the information. The remaining two lines control the source and detector qubits, serving as auxiliary qubits for state initialization and measurement of the target qubits, respectively.

The protocol, illustrated in Fig.\ref{fig:1}d, starts with all qubits at $s^\alpha(0) = 0$, such that they are effectively decoupled from each other (since $B(0) \approx 0$) and in their ground state (since $A(0)\gg k_BT$). The initialization is then realized by preparing the source qubits in a known state by forward annealing to $s^\text{source}=1$ in the presence of a polarizing bias, which is removed after the anneal completes. Each target qubit that needs an initial gate operation is strongly coupled to a unique source qubit. We then forward anneal the target qubits to $s^\text{target} = s^*$, which is fixed for the rest of the operation. Finally, we rapidly quench the source qubits back to $s^\text{source}=0$, which effectively turns off the source-target interaction. This process evolves the target qubits to their desired initial states, which are separable under the weak coupling assumption. After this transient time, the target qubits start evolving according to their own static Hamiltonian.

After the desired analog evolution time, measurement is realized by rapidly quenching the detector qubits to $s^\text{detector}=1$ and then measuring each detector qubit’s flux state, equivalently its magnetization. During this quench, each detector qubit becomes strongly coupled to its corresponding target qubit, enabling readout of the target state in a programmable basis.

The individual initial states and measurement bases of the target qubits are selected by tuning static flux biases on the corresponding source and detector qubits, respectively~\cite{SM}.  This is operationally equivalent to applying single-qubit rotation gates before and after the analog evolution, followed by measurement in the computational basis.

This protocol enables the study of the analog evolution of the target qubits under a programmable, time-independent Hamiltonian (Fig.~\ref{fig:1}d) between initialization and detection. Numerical simulations of both the state preparation and measurement processes~\cite{SM} demonstrate that these operations can be implemented with high fidelity using the device parameters and control capabilities available in current hardware. 

During the analog evolution, source and detector qubits are effectively decoupled from target qubits because $B(s) \approx 0$ when $s=s^\text{detector}=s^\text{source}=0$. When the coupling energies between target qubits are much smaller than the transverse field, it is convenient to decompose the Hamiltonian of the system comprising coupled target qubits into a dominant term and a perturbative correction: $H_\text{target}(s^*) = H_0 + H_I$ with
\begin{align}\label{eq_h0}
H_0 &= -\frac{\Delta}{2} \sum_i \sigma^z_i,
\\
H_I &= 
\sum_{\langle i,j\rangle}\frac{\mathcal{J}_{ij}}{2} \sigma^x_i \sigma^x_j
- \sum_i \frac{\delta\Delta_i}{2} \sigma^z_i,\label{eq_hi}
\end{align}
where the summation runs over target qubits only. Here, $\Delta = A(s^*)$, $\delta \Delta_i = A(s_i) - \Delta$, and $\mathcal{J}_{ij} = \sqrt{B(s_i)B(s_j)} J_{ij}$. We have also changed the basis to make $H_0$ diagonal, i.e., $\sigma^x_i = -\tau^z_i, \sigma^y_i = \tau^y_i, \sigma^z_i = \tau^x_i$. Assuming $\Delta \gg \{\mathcal{J}_{ij}, \delta \Delta_i \}$, one can now use the interaction picture and the rotating wave approximation~\cite{Kiely_TFIM_RWA_2018} (or equivalently, the secular approximation) to obtain the effective Hamiltonian 
\begin{equation}
\label{eq:H_eff}
H_\text{eff} = 
\sum_{\langle i,j\rangle} \frac{\mathcal{J}_{ij}}{4} \left(\sigma^x_i \sigma^x_j + \sigma^y_i \sigma^y_j\right)
-
\sum_i \frac{\delta\Delta_i}{2} \sigma^z_i. 
\end{equation}
This is the isotropic Heisenberg XY Hamiltonian with site-dependent longitudinal fields (hereafter referred to as the XY Hamiltonian). For the experiments that follow, $\{\delta\Delta_i\}$ are set to zero except when specified.

\section*{Experimental Results}
\subsection*{Single-Qubit Measurements}

\begin{figure}[t!]
\includegraphics[scale=1.25]{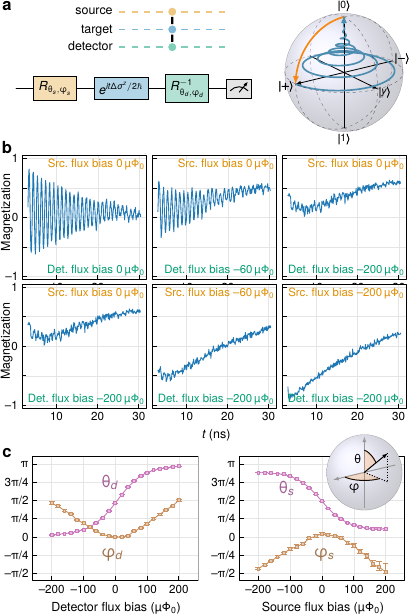}
\caption{{\bf Single-qubit Larmor precession with arbitrary-basis initialization and readout.} 
{\bf a}, Illustration of the protocol: Initial state is excited with an initial rotation gate (orange), then allowed to precess freely as it relaxes back to $\ket 0$ (blue).
{\bf b}, Measured magnetization for a single qubit for varying evolution time $t$ under the Hamiltonian $H_0$. Flux bias is applied to the detector to tilt the measurement basis (top row), interpolating from the $\sigma^x$ basis ($\theta_d=\tfrac \pi2$) to the $\sigma^z$ basis ($\theta_d\in\{0,\pi\}$) and correspondingly revealing dephasing-dominated and relaxation-dominated dynamics.  Applying flux bias to the source qubit (bottom row) rotates the initial excitation from $\ket +$ to $\ket 1$.  Short-time data ($t < \SI{2}{ns}$, omitted) reflect overlap between the initialization and detection processes.  {\bf c}, Polar angles $\theta_d$, $\theta_s$ and azimuthal angles $\varphi_d$, $\varphi_s$ extracted from single-qubit Larmor precession fits to Eq.~\eqref{eq:bloch_angles}.  Error bars indicate the 95\% confidence interval for the median angle across 124 three-qubit source-target-detector systems.}\label{fig:2}
\end{figure}

We first consider a single qubit initialized in the ground state $\ket 0$.  We excite the qubit to a state represented by the Bloch vector $\hat n_s$ parametrized by the polar and azimuthal angles $(\theta_s,\varphi_s)$: $\ket{\psi(t_0)} = R_{\theta_s,\varphi_s} \ket{0}$, where $t_0$ is the initialization time and $R_{\theta_s,\varphi_s}$ is the rotation gate corresponding to $\hat n_s$. After evolving under the Hamiltonian $H_0$ from Eq.~(\ref{eq_h0}) for time $t$, we measure the state $\ket{\psi(t)}$ along the axis given by Bloch vector $\hat n_d$. This is equivalent to applying $R^{-1}_{\theta_d,\varphi_d}$ and measuring in the computational basis. During the evolution, the qubit undergoes Larmor precession, with relaxation and dephasing represented by effective timescales $T_1$ and $T_2 = (T_1^{-1}/2 + T_\varphi^{-1})^{-1}$, respectively, where $T_\varphi$ is the pure dephasing time (Fig.~\ref{fig:2}a). Under the Lindblad master equation, the expected magnetization along $\hat n_d$ at time $t$ is given by~\cite{SM}
\begin{multline}\label{eq:bloch_angles}
\bra{\psi(t)}\sigma_{\hat n_d} \ket{\psi(t)}
= \cos\theta_d\big[1-e^{-\frac{t-t_0}{T_1}}(1{-}\cos{\theta_s})\big] 
\\ 
+ \sin \theta_d \sin \theta_s \cos\big[\Delta (t{-}t_0)/\hbar + \varphi_s {-}\varphi_d\big] e^{-\frac{t-t_0}{T_2}},
\end{multline}
where $\sigma_{\hat n_d} = R_{\theta_d,\varphi_d} \sigma^z R^{-1}_{\theta_d,\varphi_d}$ is the rotated Pauli matrix.  This allows us to experimentally relate the source- and detector-qubit flux biases to the vectors $\hat n_s$ and $\hat n_d$.

\begin{figure*}[t!]
\includegraphics[scale=1.35]{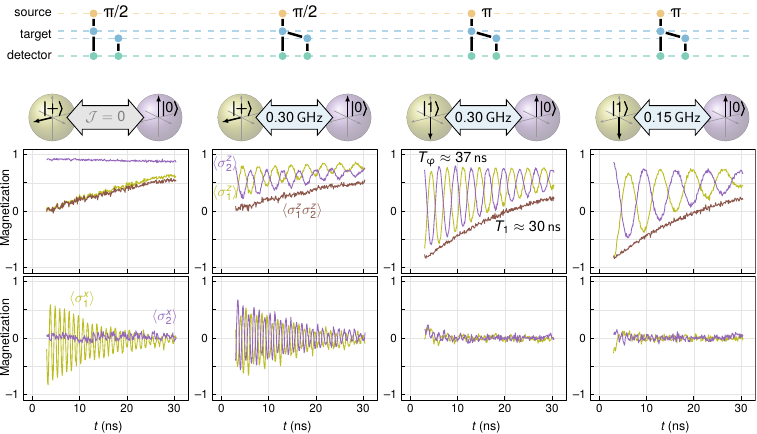}
\caption{{\bf Two-qubit spin exchange in multiple measurement bases.}
We measure a two-qubit system with target qubits $q_1$ and $q_2$ at $\Delta/h=\SI{1.0}{GHz}$. In the first two columns, the system is prepared in the product state $\ket{+0}$; in the last two columns, it is prepared in $\ket{10}$. The two-qubit exchange coupling is set to $\mathcal{J}=0$ (left), $\mathcal{J}/h=\SI{0.30}{GHz}$ (middle two columns), or $\mathcal{J}/h=\SI{0.15}{GHz}$ (right). Magnetizations are averaged over 2000 shots. Fitting the relaxation of $\langle\sigma^z\sigma^z\rangle$ and the dephasing of the oscillations in $\langle\sigma^z_1\rangle-\langle\sigma^z_2\rangle$ to an open-quantum-system model~\cite{SM} for 62 different pairs of target qubits yields median $T_1 = \SI{30}{ns}$ and $T_\varphi=\SI{37}{ns}$ for the third column.}
\label{fig:3}
\end{figure*}

With zero flux bias on the source and detector qubits, our protocol realizes an initial $\pi/2$ pulse to $\ket + = (\ket 0 + \ket 1)/\sqrt{2}$ and measurement in the $\sigma^x$ basis~\cite{SM}.  We perform the measurement with  $\Delta/h=\SI{1.0}{GHz}$ and observe the expected Larmor precession (Fig.~\ref{fig:2}b).  We then sweep the detector flux bias across a range of values from $\SI{-200}{\micro \Phi_0}$ to $\SI{200}{\micro \Phi_0}$ and extract the angles $\theta_d$ and $\varphi_d$ by fitting to the model given by~(\ref{eq:bloch_angles}) (Fig.~\ref{fig:2}c).  Median values of $T_1=\SI{32}{ns}$ and $T_\varphi=\SI{12}{ns}$ are extracted from data taken at large and zero detector flux bias, respectively. The measured $T_\varphi$ is dominated by inhomogeneous broadening due to the slow
fluctuations of $\Delta$ over different shots for averaging.  This effect will be suppressed in multi-qubit experiments due to motional narrowing~\cite{bloembergen_1948}, which scales the spread of fluctuations $W_\Delta$ to $W_\Delta^2/\mathcal{J}$~\cite{anderson_mn_1953}, increasing the observed dephasing timescale when $\mathcal{J} \gg W_\Delta$. We show in the next section that the observed dephasing timescale for multi-qubit experiments is indeed significantly longer, consistent with this effect. Assuming $T_1$, $T_\varphi$ and $\Delta$ to be constant, we extract $\theta_d$ and $\varphi_d$ across the entire range of flux bias. Then, for $\theta_d\approx 0$ (detector flux bias $\SI{-200}{\micro \Phi_0}$), we repeat a similar analysis to map out $\theta_s$ and $\varphi_s$ (Fig.~\ref{fig:2}c).

The fit demonstrates that we can independently tune $\theta_s$ and $\theta_d$ continuously between $0$ and $\pi$ by changing the source and detector flux biases, respectively.  These flux biases also determine $\varphi_s$ and $\varphi_d$.  However, the measurements are in the laboratory frame, and evolution under $H_0$ provides a time-dependent $\sigma_z$ rotation. Therefore, in the weak coupling regime, we can tune $\varphi_s$ and $\varphi_d$ independent of the polar angles by offsetting the timing of excitation and detection using anneal offsets $s_{i0}$ on the source and target qubits, respectively, allowing arbitrary initialization and measurement vectors $\hat n_s$ and $\hat n_d$ throughout the Bloch sphere.

\begin{figure*}[t!]
\includegraphics[scale=1.35]{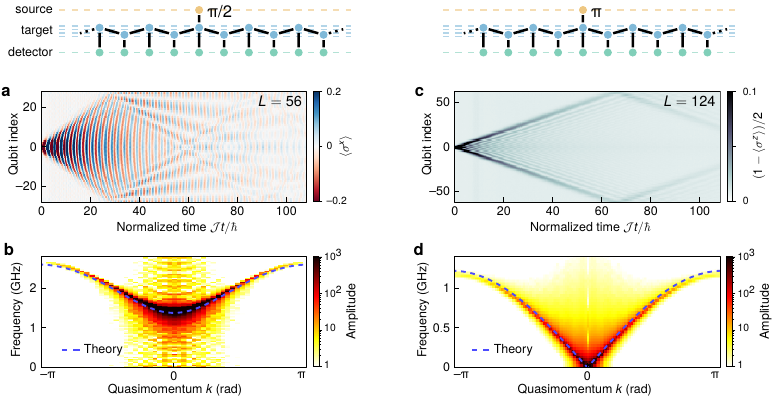}
\caption{{\bf Excitation propagation in a clean periodic one-dimensional chain.} We measure the propagation of two initial states in two bases with $\Delta/h=\SI{2.0}{GHz}$ and $\mathcal J/h=\SI{-0.60}{GHz}$.  {\bf a--b}, A single qubit is prepared in the state $\ket +$ in a chain with $L=56$ and evolved under the Hamiltonian in Eqs.~\eqref{eq_h0} and \eqref{eq_hi}, then measured in the $\sigma^x$ basis. {\bf a}, Magnetization shows ballistic propagation of information, as well as rays of destructive interference as the light cone  reaches the periodic boundary.  {\bf b}, The dispersion relation is given by the peak of the magnetization's Fourier transform, with experimental data in agreement with Eq.~\eqref{eq_dispersion}.  {\bf c--d}, A qubit is prepared in the state $\ket 1$ in a chain with $L=124$ and evolved under the Hamiltonian in Eqs.~\eqref{eq_h0} and \eqref{eq_hi}, then measured in the $\sigma^z$ basis.  {\bf c}, Again, ballistic propagation is observed throughout the chain. {\bf d}, The Fourier transform's intensity peak follows Eq.~\eqref{eq:photon}. Note that the normalized group velocity in {\bf a} and {\bf c} is unity \cite{SM}, therefore, the wavefront can propagate over distances exceeding 100 qubits before coherence is lost, which can cover the entire processor.} 
\label{fig:4}
\end{figure*}

In the experiments that follow, we mitigate the effect of state preparation and measurement (SPAM) errors by repeating each experiment for both signs of the polarizing field applied to the source qubit, which transforms the initial state as $\theta_s \rightarrow \pi - \theta_s$, $\varphi_s \rightarrow \varphi_s + \pi$, and by repeating the measurement at $\theta_d$ and $\pi - \theta_d$ while holding $\varphi_d$ fixed,  implemented using equal and opposite flux offsets on the detector.

\subsection*{Multi-Qubit Experiments}

Having demonstrated analog–digital computation on a single-qubit system, we now turn to multi-qubit operations. Our goal is to show, through agreement with theory in the weak-coupling regime, that reliable initialization and measurement remain possible even when the static coupling terms remain on during initialization and readout.

In a system of two coupled target qubits with $\delta\Delta_i=0$, the effective Hamiltonian is (recall Eq. \eqref{eq:H_eff})
\begin{equation}\label{eq:H_2Q}
H_\text{2Q} = \frac{\mathcal{J}}{4} \left(\sigma_1^x\sigma_2^x + \sigma_1^y\sigma_2^y\right).
\end{equation}
This Hamiltonian drives a spin exchange between the two qubits. Fig.~\ref{fig:3} shows that we can see this behavior in both the $\sigma^x$ and $\sigma^z$ bases of the two qubits.  As in the single-qubit case, we can derive analytical expressions for the evolution under $H_\text{2Q}$ assuming independent noise on each qubit with identical relaxation and dephasing timescales $T_1$ and $T_\varphi$~\cite{SM}.  A fit of the data to the analytical solutions gives excellent agreement, showing that in this weak-coupling regime the measurements are well described by independent single-qubit operators despite the always-on coupling. The median extracted dephasing time $T_\varphi = \SI{37}{ns}$ is much longer than the value extracted from the single-qubit data fit to \eqref{eq:bloch_angles}, showing that the latter was limited by inhomogeneous broadening of $\Delta$. The two-qubit evolution is less sensitive to the inhomogeneous broadening seen for single-qubit systems but will still be limited by a combination of slow fluctuations of $\mathcal{J}$ and $\Delta$. Therefore, $T_\varphi = \SI{37}{ns}$ only gives a lower bound on the actual dephasing timescale per shot, which would limit coherent evolution of multi-qubit systems. Notably, experiments on real materials containing single and two-qubit  subsystems reveal a similar enhancement of coherence for qubit pairs relative to single qubits \cite{Beckert24}. 

We can now extend our analog-digital protocol to study evolution under a many-body Hamiltonian.  First, we embed a 56-qubit, periodic 1D chain of target qubits, each with a dedicated source and detector qubit on the 1178-qubit processor, with all qubits set to $\Delta/h=\SI{2.0}{GHz}$, $\delta\Delta_i = 0$, and nearest-neighbor coupling $\mathcal{J}/h = \SI{-0.60}{GHz}$. We prepare a single qubit in the chain in the $\ket +$ state and observe the system's time evolution in the $\sigma^x$ basis. Fig.~\ref{fig:4}a shows the coherent propagation of the excitation around the closed chain, and the resulting interference pattern as the propagating
fronts pass through each other on the opposite side
of the ring. This quantity is proportional to the ground-state retarded Green's function $G_{xx}^R(x,t)$, whose two-dimensional Fourier transform yields the dispersion relation (see Supplementary Material).

This system can be mapped to a non-interacting spinless fermionic model with the Jordan–Wigner transformation \cite{jordan1928paulische, Mbengquantum2024, Dziamaraga-exact-solution}. For a periodic chain, the single-particle dispersion of the corresponding Bogoliubov–de Gennes Hamiltonian is
\begin{equation}
E(k) = \pm \sqrt{\left(\Delta + \mathcal{J}\cos k\right)^2 + \mathcal{J}^2\sin^2 k},
\label{eq_dispersion}
\end{equation}
where $k$ is the wave vector within the first Brillouin zone. By performing a two-dimensional Fourier transform of the space–time experimental data, we obtain the sharp
single-paramagnon dispersion shown in Fig.~\ref{fig:4}b (only positive frequencies are shown), in excellent agreement with the model prediction in Eq.~(\ref{eq_dispersion}) with no free fit parameters.

\begin{figure}[t!]
\includegraphics[scale=1.45]{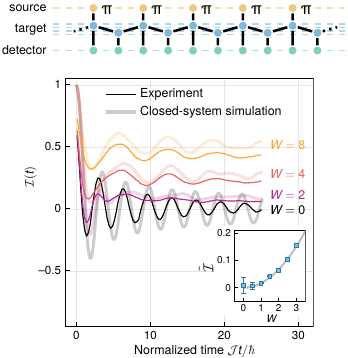}
\caption{{\bf Anderson localization in a disordered chain}. Every other qubit is prepared in $\ket 1$ in a periodic chain ($L{=124}$) with disorder $W$ applied through qubit detunings $\delta\Delta_i$ uniformly distributed in $[-\mathcal J\tfrac W 2, \mathcal J\tfrac W 2 ]$, where $\mathcal J/h=\SI{0.20}{GHz}$ and the quantizing field is $\Delta/h=\SI{2.0}{GHz}$. In the clean chain ($W=0$), the imbalance oscillates about zero.  As disorder is increased, a finite imbalance persists, consistent with localization. Experimental and numerical data are averaged over 20 disorder realizations. Inset: For small $W$, averaged QPU imbalance between normalized times $\mathcal{J}t/\hbar=6\pi$ and $8\pi$ scales quadratically \cite{ros_remanent_2017} as $\tilde{\mathcal I} = 0.017\,W^{2}$ (gray curve). Error bars indicate uncertainty over time.}
\label{fig:5}
\end{figure}

We next initialize a single qubit in the chain to $\ket 1$ with a $\pi$ pulse and observe the spread of the excitation energy in time by measuring the qubits in the $\sigma^z$ basis (Fig.~\ref{fig:4}c).  As the
model calculation~\cite{SM} shows, in Fourier space this results in a continuum with sharp peaks at
\begin{equation}
f(k) = \pm 2\mathcal J\sin(k/2)/h,
\label{eq:photon}
\end{equation}
which is indeed reflected in the experimental measurements in Fig.~\ref{fig:4}d.

To further demonstrate fine control over the effective XY Hamiltonian~\eqref{eq:H_eff}, we explore Anderson localization in a periodic 1D chain by programming individual detunings $\delta\Delta_i$ on the target qubits. Anderson localization and its many-body generalization describe how disorder halts the propagation of quantum particles and excitations through destructive interference and inefficient tunneling, resulting in their localization~\cite{PhysRev.109.1492, basko2006metal, nandkishore2015many, RevModPhys.91.021001}. In one dimension, arbitrarily weak disorder localizes all single-particle states.

To probe this phenomenon, we excite odd-indexed qubits, initializing the system in the staggered configuration $\ket{1010\ldots}$.  In a delocalized system, excitations should spread through the chain and erase
the memory of the initial staggered state. The persistence of the initial spin density wave can be quantified by the imbalance~\cite{schreiber2015observation}, defined as
\begin{align}
\mathcal I(t) = \frac{p_{\rm odd}(t) - p_{\rm even}(t)}{p_{\rm odd}(t) + p_{\rm even}(t)},
\end{align}
where $p_\text{odd}$ and $p_\text{even}$ are average excitation probabilities of the odd and even qubits, respectively, with
\begin{align}
p_{\rm odd} &= \frac{1}{L} \sum_{i=1}^{L/2} \Big(1 - \langle\sigma^z_{2i-1}\rangle
\Big),\\
p_{\rm even} &= \frac{1}{L} \sum_{i=1}^{L/2} \Big(1 - \langle\sigma^z_{2i}\rangle
\Big).
\end{align}
The initial state corresponds to a maximum imbalance, $\mathcal I(0) = 1$. In a delocalized clean chain, $\mathcal I(t)$ is expected to decay toward zero, up to finite-size recurrences.

Starting from a clean chain, we introduce disorder $W$ by adding uniformly distributed detunings $\delta\Delta_i$ in the interval $[-\mathcal J\tfrac W2, \mathcal J\tfrac W2]$.  Results are shown in Fig.~\ref{fig:5} for varying $W$.  As expected, the clean chain ($W=0$) is delocalized, with the imbalance decaying and oscillating about zero. Increasing disorder leads to a finite long-time imbalance consistent with Anderson localization over the finite system and time window studied. These results are in close agreement with closed-system simulations; the remaining deviations are likely due to decoherence and readout nonidealities, while uniform excitation loss would mostly cancel in the normalized imbalance.  

The inset of Fig.~\ref{fig:5} shows average imbalance $\tilde{\mathcal I}$ for $\mathcal{J}t/\hbar \in [6\pi, 8\pi]$. We observe, for the first time to our knowledge, a quadratic dependence on $W$, in agreement with the theoretical prediction \cite{ros_remanent_2017}.

\section*{Conclusions}

We have demonstrated that commercially available QA processors can be used to perform analog-digital quantum computation using multicolor annealing protocols in which different subsets of qubits follow independent anneal schedules. Operationally, this is equivalent to implementing single-qubit rotations on either side of analog evolution under a programmable many-body Hamiltonian that, in the weak-coupling regime, is well described by an effective XY Hamiltonian. Despite always-on couplings during the excitation and readout processes, our multi-qubit results in the weak-coupling regime show excellent agreement with theory. The breadth of capabilities shown here, including arbitrary-basis initialization and readout, demonstrates that quantum annealers can simulate dynamics beyond the constraints of the standard annealing protocol. During the preparation of this manuscript, two other works \cite{geim_engineering_2026,lee_2026} were published, demonstrating, among other things, quantum walks in XY chains on 31 qubits \cite{geim_engineering_2026} and 50 qubits \cite{lee_2026}, respectively. This provides an opportunity to directly compare the performance of quantum simulators in different modalities.

This work opens many avenues for quantum simulation on quantum annealers, in particular to study the dynamics of non-integrable, higher-dimensional interacting models, and the interplay of disorder and interactions. Individual single-qubit rotations and measurements also enable Green's function extraction via many-qubit interferometry experiments. Other experiments may include combining quantum annealing with digital operations to perform mid-anneal excitation and readout in arbitrary bases.

These first demonstrations of ADQC in a quantum annealer used a QPU already in a production environment, not one designed with analog-digital quantum computation in mind. Future processors will benefit from design choices to minimize overhead and maximize performance of ADQC protocols, allowing the full use of the complex connectivity in annealing QPUs without sacrificing qubits to auxiliary roles.\\

\section*{Acknowledgments}
\vspace{-1mm}

We gratefully acknowledge helpful discussions with Anatoly Smirnov,
Vladimir Vargas-Calderón, Javier Toledo Marin, Humberto Munoz Bauza, and contributions from the D-Wave team in calibrating the processor used in this work, as well as in developing software infrastructure to enable multicolor annealing experiments through the Python-based API. M.Kh. thanks Igor Herbut and acknowledges partial support from Mitacs. For D.A.L., this material is based upon work supported by, or in part by, the U. S. Army Research Laboratory and the U. S. Army Research Office under contract/grant number W911NF2310255. G.A. was supported by the European Research Council under the European Union’s Horizon 2020 research and innovation program HERO (grant agreement no. 810451).

\bibliography{adqc}

@article{Beckert24,
  title={Emergence of highly coherent two-level systems in a noisy and dense quantum network},
  author={Beckert, Adrian and Grimm, Manuel and Wili, Nino and Tschaggelar, Ren{\'e} and Jeschke, Gunnar and Matmon, Guy and Gerber, Simon and M{\"u}ller, Markus and Aeppli, Gabriel},
  journal={Nature Physics},
  volume={20},
  number={3},
  pages={472--478},
  year={2024},
  publisher={Nature Publishing Group UK London},
  url = {https://doi.org/10.1038/s41567-023-02321-y}
}

@misc{lee_2026,
      title={Benchmarking quantum simulation with neutron-scattering experiments}, 
      author={Yi-Ting Lee and Keerthi Kumaran and Bibek Pokharel and Allen Scheie and Colin L. Sarkis and David A. Tennant and Travis Humble and André Schleife and Abhinav Kandala and Arnab Banerjee},
      year={2026},
      eprint={2603.15608},
      archivePrefix={arXiv},
      url={https://arxiv.org/abs/2603.15608}, 
}

@misc{imoto_2025,
      title={Universal quantum computation using quantum annealing with the transverse-field Ising Hamiltonian}, 
      author={Takashi Imoto and Yuki Susa and Ryoji Miyazaki and Yuichiro Matsuzaki},
      year={2025},
      eprint={2402.19114},
      archivePrefix={arXiv},
      primaryClass={quant-ph},
      url={https://arxiv.org/abs/2402.19114}, 
}

@article{bloembergen_1948,
  title = {Relaxation Effects in Nuclear Magnetic Resonance Absorption},
  author = {Bloembergen, N. and Purcell, E. M. and Pound, R. V.},
  journal = {Phys. Rev.},
  volume = {73},
  issue = {7},
  pages = {679--712},
  numpages = {0},
  year = {1948},
  month = {Apr},
  publisher = {American Physical Society},
  doi = {10.1103/PhysRev.73.679},
  url = {https://link.aps.org/doi/10.1103/PhysRev.73.679}
}

@article{anderson_mn_1953,
  title = {Exchange Narrowing in Paramagnetic Resonance},
  author = {Anderson, P. W. and Weiss, P. R.},
  journal = {Rev. Mod. Phys.},
  volume = {25},
  issue = {1},
  pages = {269--276},
  numpages = {0},
  year = {1953},
  month = {Jan},
  publisher = {American Physical Society},
  doi = {10.1103/RevModPhys.25.269},
  url = {https://link.aps.org/doi/10.1103/RevModPhys.25.269}
}

@article{basko2006metal,
  title={Metal--insulator transition in a weakly interacting many-electron system with localized single-particle states},
  author={Basko, Denis M and Aleiner, Igor L and Altshuler, Boris L},
  journal={Annals of physics},
  volume={321},
  number={5},
  pages={1126--1205},
  year={2006},
  publisher={Elsevier},
  url ={https://www.sciencedirect.com/science/article/abs/pii/S0003491605002630}
}

@article{RevModPhys.91.021001,
  title = {Colloquium: Many-body localization, thermalization, and entanglement},
  author = {Abanin, Dmitry A. and Altman, Ehud and Bloch, Immanuel and Serbyn, Maksym},
  journal = {Rev. Mod. Phys.},
  volume = {91},
  issue = {2},
  pages = {021001},
  numpages = {26},
  year = {2019},
  month = {May},
  publisher = {American Physical Society},
  doi = {10.1103/RevModPhys.91.021001},
  url = {https://link.aps.org/doi/10.1103/RevModPhys.91.021001}
}

@article{nandkishore2015many,
  title={Many-body localization and thermalization in quantum statistical mechanics},
  author={Nandkishore, Rahul and Huse, David A},
  journal={Annu. Rev. Condens. Matter Phys.},
  volume={6},
  number={1},
  pages={15--38},
  year={2015},
  publisher={Annual Reviews},
  url = {https://www.annualreviews.org/content/journals/10.1146/annurev-conmatphys-031214-014726}
}

@misc{SM,
title={See Supplementary Materials},
}

@article{PhysRev.117.648,
  title = {Quasi-Particles and Gauge Invariance in the Theory of Superconductivity},
  author = {Nambu, Yoichiro},
  journal = {Phys. Rev.},
  volume = {117},
  issue = {3},
  pages = {648--663},
  numpages = {0},
  year = {1960},
  month = {Feb},
  publisher = {American Physical Society},
  doi = {10.1103/PhysRev.117.648},
  url = {https://link.aps.org/doi/10.1103/PhysRev.117.648}
}

@Article{CCJJqubit,
  title = {Experimental Demonstration of a Robust and Scalable Flux Qubit},
  year = 2010,
  month = apr,
  journal = {Physical Review B},
  volume = {81},
  number = {13},
  pages = {134510},
  doi = {10.1103/PhysRevB.81.134510},
  author = {Harris, Richard and Johansson, J. and Berkley, A. J. and Johnson, M. W. and Lanting, T. and others}
}

@Article{CJJcoupler,
  title = {Compound {{Josephson-junction}} Coupler for Flux Qubits with Minimal Crosstalk},
  author = {Harris, R. and Lanting, T. and Berkley, A. J. and Johansson, J. and Johnson, M. W. and Bunyk, P. and Ladizinsky, E. and Ladizinsky, N. and Oh, T. and Han, S.},
  year = 2009,
  month = aug,
  journal = {Physical Review B},
  volume = {80},
  number = {5},
  pages = {052506},
  doi = {10.1103/PhysRevB.80.052506},
  copyright = {http://link.aps.org/licenses/aps-default-license}
}

@article{Poletto_2009,
    doi = {10.1088/1367-2630/11/1/013009},
    url = {https://doi.org/10.1088/1367-2630/11/1/013009},
    year = {2009},
    month = {jan},
    publisher = {},
    volume = {11},
    number = {1},
    pages = {013009},
    author = {Poletto, S and Chiarello, F and Castellano, M G and Lisenfeld, J and Lukashenko, A and Cosmelli, C and Torrioli, G and Carelli, P and Ustinov, A V},
    title = {Coherent oscillations in a superconducting tunable flux qubit manipulated without microwaves},
    journal = {New Journal of Physics}
    }

@article{earlyfluxqubit,
    author = {I. Chiorescu  and Y. Nakamura  and C. J. P. M. Harmans  and J. E. Mooij },
    title = {Coherent Quantum Dynamics of a Superconducting Flux Qubit},
    journal = {Science},
    volume = {299},
    number = {5614},
    pages = {1869-1871},
    year = {2003},
    doi = {10.1126/science.1081045},
    URL = {https://www.science.org/doi/abs/10.1126/science.1081045}
    }

@article{schreiber2015observation,
  title = {Observation of Many-Body Localization of Interacting Fermions in a Quasirandom Optical Lattice},
  author = {Schreiber, Michael and Hodgman, Sean S. and Bordia, Pranjal and L{\"u}schen, Henrik P. and Fischer, Mark H. and Vosk, Ronen and Altman, Ehud and Schneider, Ulrich and Bloch, Immanuel},
  year = 2015,
  month = aug,
  journal = {Science},
  volume = {349},
  number = {6250},
  pages = {842--845},
  doi = {10.1126/science.aaa7432},
  copyright = {http://www.sciencemag.org/about/science-licenses-journal-article-reuse}
}

@article{Lanting_2017_nonuniform_driver,
  title = {Experimental demonstration of perturbative anticrossing mitigation using nonuniform driver Hamiltonians},
  author = {Lanting, Trevor and King, Andrew D. and Evert, Bram and Hoskinson, Emile},
  journal = {Phys. Rev. A},
  volume = {96},
  issue = {4},
  pages = {042322},
  numpages = {8},
  year = {2017},
  month = {Oct},
  publisher = {American Physical Society},
  doi = {10.1103/PhysRevA.96.042322},
  url = {https://link.aps.org/doi/10.1103/PhysRevA.96.042322}
}

@article{PhysRev.109.1492,
  title = {Absence of Diffusion in Certain Random Lattices},
  author = {Anderson, P. W.},
  journal = {Phys. Rev.},
  volume = {109},
  issue = {5},
  pages = {1492--1505},
  numpages = {0},
  year = {1958},
  month = {Mar},
  publisher = {American Physical Society},
  doi = {10.1103/PhysRev.109.1492},
  url = {https://link.aps.org/doi/10.1103/PhysRev.109.1492}
}

@article{Lidar_notes_2019,
  title={Lecture notes on the theory of open quantum systems},
  author={Lidar, Daniel A},
  journal={arXiv:1902.00967},
  year={2019},
  url = {https://arxiv.org/abs/1902.00967}
}

@article{Kiely_TFIM_RWA_2018,
  title = {Relationship between the Transverse-Field {{Ising}} Model and the {XY} Model via the Rotating-Wave Approximation},
  author = {Kiely, Thomas G. and Freericks, J. K.},
  year = 2018,
  month = feb,
  journal = {Physical Review A},
  volume = {97},
  number = {2},
  pages = {023611},
  doi = {10.1103/PhysRevA.97.023611}
}

@article{jordan1928paulische,
  title = {\"Uber Das {{Paulische \"Aquivalenzverbot}}},
  author = {Jordan, P. and Wigner, E.},
  year = 1928,
  month = sep,
  journal = {Zeitschrift f\"ur Physik},
  volume = {47},
  number = {9-10},
  pages = {631--651},
  doi = {10.1007/BF01331938}
}

@article{barends_digitized_2016,
  title = {Digitized Adiabatic Quantum Computing with a Superconducting Circuit},
  year = 2016,
  journal = {Nature},
  volume = {534},
  number = {7606},
  pages = {222--226},
  publisher = {Nature Publishing Group},
  doi = {10.1038/nature17658},
  author = {Barends, R. and Shabani, A. and Lamata, L. and Kelly, J. and Mezzacapo, A. and others}
}

@article{bluvstein_quantum_2022,
  title = {A Quantum Processor Based on Coherent Transport of Entangled Atom Arrays},
  year = 2022,
  month = apr,
  journal = {Nature},
  volume = {604},
  number = {7906},
  pages = {451--456},
  doi = {10.1038/s41586-022-04592-6},
  author = {Bluvstein, Dolev and Levine, Harry and Semeghini, Giulia and Wang, Tout T. and Ebadi, Sepehr and others}
}

@article{Boixo2014a,
  title = {Evidence for Quantum Annealing with More than One Hundred Qubits},
  author = {Boixo, Sergio and R{\o}nnow, Troels F. and Isakov, Sergei V. and Wang, Zhihui and Wecker, David and Lidar, Daniel A. and Martinis, John M. and Troyer, Matthias},
  year = 2014,
  month = mar,
  journal = {Nature Physics},
  volume = {10},
  number = {3},
  pages = {218--224},
  doi = {10.1038/nphys2900}
}

@article{Bunyk2014,
  title = {Architectural {{Considerations}} in the {{Design}} of a {{Superconducting Quantum Annealing Processor}}},
  year = 2014,
  month = aug,
  journal = {IEEE Transactions on Applied Superconductivity},
  volume = {24},
  number = {4},
  pages = {1--10},
  doi = {10.1109/TASC.2014.2318294},
  author = {Bunyk, Paul I and Hoskinson, Emile M. and Johnson, Mark W. and Tolkacheva, Elena and Altomare, Fabio and others}
}

@article{ebadi_quantum_2021,
  title = {Quantum Phases of Matter on a 256-Atom Programmable Quantum Simulator},
  year = 2021,
  month = jul,
  journal = {Nature},
  volume = {595},
  number = {7866},
  pages = {227--232},
  doi = {10.1038/s41586-021-03582-4},
  author = {Ebadi, Sepehr and Wang, Tout T. and Levine, Harry and Keesling, Alexander and Semeghini, Giulia and others}
}

@article{haghshenas_digital_2025,
  title={Digital quantum magnetism at the frontier of classical simulations},
  author={Haghshenas, Reza and Chertkov, Eli and Mills, Michael and Kadow, Wilhelm and Lin, Sheng-Hsuan and Chen, Yi-Hsiang and Cade, Chris and Niesen, Ido and Begu{\v{s}}i{\'c}, Tomislav and Rudolph, Manuel S and others},
  journal={arXiv:2503.20870},
  year={2025},
  url = {https://arxiv.org/abs/2503.20870}
}

@article{harris_experimental_2010,
  title = {Experimental Investigation of an Eight-Qubit Unit Cell in a Superconducting Optimization Processor},
  year = 2010,
  month = jul,
  journal = {Physical Review B},
  volume = {82},
  number = {2},
  pages = {024511},
  doi = {10.1103/PhysRevB.82.024511},
  copyright = {http://link.aps.org/licenses/aps-default-license},
  author = {Harris, R. and Johnson, M. W. and Lanting, T. and Berkley, A. J. and Johansson, J. and others}
}

@article{harris_phase_2018,
  title = {Phase Transitions in a Programmable Quantum Spin Glass Simulator},
  year = 2018,
  month = jul,
  journal = {Science},
  volume = {361},
  number = {6398},
  pages = {162--165},
  doi = {10.1126/science.aat2025},
  author = {Harris, Richard and Sato, Yuki and Berkley, Andrew J. and Reis, M and Altomare, Fabio and others}
}

@article{johnson_quantum_2011,
  title = {Quantum Annealing with Manufactured Spins},
  year = 2011,
  month = may,
  journal = {Nature},
  volume = {473},
  number = {7346},
  pages = {194--198},
  doi = {10.1038/nature10012},
  author = {Johnson, Mark W. and Amin, Mohammad H. and Gildert, S. and Lanting, Trevor and Hamze, Firas and others}
}

@article{kim_evidence_2023,
  title = {Evidence for the Utility of Quantum Computing before Fault Tolerance},
  year = 2023,
  month = jun,
  journal = {Nature},
  volume = {618},
  number = {7965},
  pages = {500--505},
  publisher = {Springer US},
  doi = {10.1038/s41586-023-06096-3},
  author = {Kim, Youngseok and Eddins, Andrew and Anand, Sajant and Wei, Ken Xuan and {\noopsort{berg}}{van den Berg}, Ewout and others}
}

@article{king_beyondclassical_2025,
  title = {Beyond-Classical Computation in Quantum Simulation},
  year = 2025,
  month = apr,
  journal = {Science},
  volume = {388},
  number = {6743},
  pages = {199--204},
  doi = {10.1126/science.ado6285},
  author = {King, Andrew D. and Nocera, Alberto and Rams, Marek M. and Dziarmaga, Jacek and Wiersema, Roeland and others}
}

@article{king_observation_2018,
  title = {Observation of Topological Phenomena in a Programmable Lattice of 1,800 Qubits},
  year = 2018,
  journal = {Nature},
  volume = {560},
  number = {7719},
  pages = {456--460},
  doi = {10.1038/s41586-018-0410-x},
  author = {King, Andrew D. and Carrasquilla, Juan and Raymond, Jack and Ozfidan, Isil and Andriyash, Evgeny and others}
}

@article{king_quantum_2023,
  title = {Quantum Critical Dynamics in a 5,000-Qubit Programmable Spin Glass},
  year = 2023,
  month = may,
  journal = {Nature},
  volume = {617},
  number = {7959},
  pages = {61--66},
  doi = {10.1038/s41586-023-05867-2},
  author = {King, Andrew D. and Raymond, Jack and Lanting, Trevor and Harris, Richard and Zucca, Alex and others}
}

@article{lidar_digitalanalogdigital_2025,
  title={Digital-Analog-Digital Quantum Supremacy},
  author={Lidar, Daniel},
  journal={arXiv:2512.07127},
  year={2025},
  url = {https://arxiv.org/abs/2512.07127}
}

@article{lunkin_evidence_2026,
  title={Evidence for a two-dimensional quantum glass state at high temperatures},
  author={Lunkin, Aleksey and Ticea, Nicole S and Kumar, Shashwat and Miao, Connie and Choi, Jaehong and Alghadeer, Mohammed and Drozdov, Ilya and Abanin, Dmitry and Abbas, Amira and Acharya, Rajeev and others},
  journal={arXiv:2601.01309},
  year={2026},
  url = {https://arxiv.org/abs/2601.01309}
}

@article{manovitz_quantum_2025,
  title = {Quantum Coarsening and Collective Dynamics on a Programmable Simulator},
  year = 2025,
  month = feb,
  journal = {Nature},
  volume = {638},
  number = {8049},
  pages = {86--92},
  doi = {10.1038/s41586-024-08353-5},
  author = {Manovitz, Tom and Li, Sophie H. and Ebadi, Sepehr and Samajdar, Rhine and Geim, Alexandra A. and others}
}

@article{miessen_benchmarking_2024,
  title = {Benchmarking {{Digital Quantum Simulations Above Hundreds}} of {{Qubits Using Quantum Critical Dynamics}}},
  author = {Miessen, Alexander and Egger, Daniel J. and Tavernelli, Ivano and Mazzola, Guglielmo},
  year = 2024,
  month = nov,
  journal = {PRX Quantum},
  volume = {5},
  number = {4},
  pages = {040320},
  doi = {10.1103/PRXQuantum.5.040320}
}

@article{scholl_quantum_2021,
  title = {Quantum Simulation of {{2D}} Antiferromagnets with Hundreds of {{Rydberg}} Atoms},
  year = 2021,
  month = jul,
  journal = {Nature},
  volume = {595},
  number = {7866},
  pages = {233--238},
  doi = {10.1038/s41586-021-03585-1},
  author = {Scholl, Pascal and Schuler, Michael and Williams, Hannah J. and Eberharter, Alexander A. and Barredo, Daniel and others}
}

@article{shaw_benchmarking_2024,
  title = {Benchmarking Highly Entangled States on a 60-Atom Analogue Quantum Simulator},
  author = {Shaw, Adam L. and Chen, Zhuo and Choi, Joonhee and Mark, Daniel K. and Scholl, Pascal and Finkelstein, Ran and Elben, Andreas and Choi, Soonwon and Endres, Manuel},
  year = 2024,
  month = apr,
  journal = {Nature},
  volume = {628},
  number = {8006},
  pages = {71--77},
  doi = {10.1038/s41586-024-07173-x}
}

@article{zhang_observation_2017,
  title = {Observation of a Many-Body Dynamical Phase Transition with a 53-Qubit Quantum Simulator},
  author = {Zhang, J and Pagano, G and Hess, P W and Kyprianidis, A and Becker, P and Kaplan, H and Gorshkov, A V and Gong, Z.-X. and Monroe, C.},
  year = 2017,
  month = nov,
  journal = {Nature},
  volume = {551},
  number = {7682},
  pages = {601--604},
  publisher = {Nature Publishing Group},
  doi = {10.1038/nature24654}
}

@article{Parra-RodriguezDigitalanalog2020,
  title = {Digital-Analog Quantum Computation},
  author = {{Parra-Rodriguez}, Adrian and Lougovski, Pavel and Lamata, Lucas and Solano, Enrique and Sanz, Mikel},
  year = 2020,
  month = feb,
  journal = {Physical Review A},
  volume = {101},
  number = {2},
  pages = {022305},
  publisher = {American Physical Society},
  doi = {10.1103/PhysRevA.101.022305},
  url = {https://link.aps.org/doi/10.1103/PhysRevA.101.022305},
  urldate = {2026-02-05}
}

@article{geim_engineering_2026,
  title={Engineering quantum criticality and dynamics on an analog-digital simulator},
  author={Geim, Alexandra A and Koyluoglu, Nazli Ugur and Evered, Simon J and Sahay, Rahul and Li, Sophie H and Xu, Muqing and Bluvstein, Dolev and Gjonbalaj, Nik O and Maskara, Nishad and Kalinowski, Marcin and others},
  journal={arXiv:2602.18555},
  year={2026},
  url = {https://arxiv.org/abs/2602.18555}
}

@article{KumarDigitalanalog2025,
  title = {Digital-Analog Quantum Computing of Fermion-Boson Models in Superconducting Circuits},
  author = {Kumar, Shubham and Hegade, Narendra N. and Visuri, Anne-Maria and Bhargava, Balaganchi A. and Hernandez, Juan F. R. and Solano, E. and {Albarr{\'a}n-Arriagada}, F. and Barrios, G. Alvarado},
  year = 2025,
  month = mar,
  journal = {npj Quantum Information},
  volume = {11},
  number = {1},
  pages = {43},
  publisher = {Nature Publishing Group},
  issn = {2056-6387},
  doi = {10.1038/s41534-025-01001-4},
  url = {https://www.nature.com/articles/s41534-025-01001-4},
  urldate = {2026-02-05},
  langid = {english}
}

@article{KatzHybrid2025,
  title={Hybrid digital-analog protocols for simulating quantum multi-body interactions},
  author={Katz, Or and Schuckert, Alexander and Wang, Tianyi and Crane, Eleanor and Gorshkov, Alexey V and Cetina, Marko},
  journal={arXiv:2512.21385},
  year={2025},
  url={https://arxiv.org/abs/2512.21385}
}

@article{AndersenThermalization2025a,
  title = {Thermalization and Criticality on an Analogue--Digital Quantum Simulator},
  year = 2025,
  month = feb,
  journal = {Nature},
  volume = {638},
  number = {8049},
  pages = {79--85},
  doi = {10.1038/s41586-024-08460-3},
  author = {Andersen, T. I. and Astrakhantsev, N. and Karamlou, A. H. and Berndtsson, J. and Motruk, J. and others}
}

@article{Mbengquantum2024,
  title = {The Quantum {{Ising}} Chain for Beginners},
  author = {Mbeng, Glen Bigan and Russomanno, Angelo and Santoro, Giuseppe E.},
  year = 2024,
  month = jun,
  journal = {SciPost Physics Lecture Notes},
  pages = {82},
  issn = {2590-1990},
  doi = {10.21468/SciPostPhysLectNotes.82},
  url = {https://scipost.org/10.21468/SciPostPhysLectNotes.82},
  urldate = {2024-08-24},
  langid = {english}
}

@book{Sachdev_2011, 
    place={Cambridge},
    edition={2},
    title={Quantum Phase Transitions}, publisher={Cambridge University Press}, author={Sachdev, Subir},
    year={2011}
}

@article{Cirac-MPSreview,
  title = {Matrix product states and projected entangled pair states: Concepts, symmetries, theorems},
  author = {Cirac, J. Ignacio and P\'erez-Garc\'{\i}a, David and Schuch, Norbert and Verstraete, Frank},
  journal = {Rev. Mod. Phys.},
  volume = {93},
  issue = {4},
  pages = {045003},
  numpages = {65},
  year = {2021},
  month = {Dec},
  publisher = {American Physical Society},
  doi = {10.1103/RevModPhys.93.045003},
  url = {https://link.aps.org/doi/10.1103/RevModPhys.93.045003}
}

@Article{YASTN,
      title={{YASTN: Yet another symmetric tensor networks; A Python library for Abelian symmetric tensor network calculations}},
      author={Marek M. Rams and Gabriela Wójtowicz and Aritra Sinha and Juraj Hasik},
      journal={SciPost Phys. Codebases},
      pages={52},
      year={2025},
      publisher={SciPost},
      doi={10.21468/SciPostPhysCodeb.52},
      url={https://scipost.org/10.21468/SciPostPhysCodeb.52},
}

@article{Dziamaraga-exact-solution,
  title = {{Dynamics of a Quantum Phase Transition: Exact Solution of the Quantum Ising Model}},
  author = {Dziarmaga, Jacek},
  journal = {Phys. Rev. Lett.},
  volume = {95},
  issue = {24},
  pages = {245701},
  numpages = {4},
  year = {2005},
  month = {Dec},
  publisher = {American Physical Society},
  doi = {10.1103/PhysRevLett.95.245701},
  url = {https://link.aps.org/doi/10.1103/PhysRevLett.95.245701}
}

@misc{dwavequantumExperimentalResearch,
	author = {},
	title = {{E}xperimental {R}esearch -- {D}-{W}ave {Q}uantum {C}omputing {P}roducts documentation --- docs.dwavequantum.com},
	howpublished = {\url{https://docs.dwavequantum.com/en/latest/quantum\_research/experimental\_research.html\#multicolor-annealing}},
	year = {2026},
	note = {[Accessed 12-03-2026]},
}

@misc{dwavequantumOcean,
	author = {},
	title = {{O}cean {S}{D}{K} -- {D}-{W}ave {Q}uantum {C}omputing {P}roducts documentation --- docs.dwavequantum.com},
	howpublished = {\url{https://docs.dwavequantum.com/en/latest/ocean/index.html}},
	year = {2026},
	note = {[Accessed 12-03-2026]},
}

@article{ros_remanent_2017,
  title = {Remanent {{Magnetization}}: {{Signature}} of {{Many-Body Localization}} in {{Quantum Antiferromagnets}}},
  shorttitle = {Remanent {{Magnetization}}},
  author = {Ros, V. and M{\"u}ller, M.},
  year = 2017,
  month = jun,
  journal = {Physical Review Letters},
  volume = {118},
  number = {23},
  pages = {237202},
  doi = {10.1103/PhysRevLett.118.237202},
  copyright = {http://link.aps.org/licenses/aps-default-license}
}

@article{farhi_quantum_2001,
  title = {A {{Quantum Adiabatic Evolution Algorithm Applied}} to {{Random Instances}} of an {{NP-Complete Problem}}},
  author = {Farhi, Edward and Goldstone, Jeffrey and Gutmann, Sam and Lapan, Joshua and Lundgren, Andrew and Preda, Daniel},
  year = 2001,
  month = apr,
  journal = {Science},
  volume = {292},
  number = {5516},
  pages = {472--475},
  doi = {10.1126/science.1057726}
}

@article{albash_adiabatic_2018,
  title = {Adiabatic Quantum Computation},
  author = {Albash, Tameem and Lidar, Daniel A.},
  year = 2018,
  month = jan,
  journal = {Reviews of Modern Physics},
  volume = {90},
  number = {1},
  pages = {015002},
  doi = {10.1103/RevModPhys.90.015002}
}

@article{KnapProbing2013,
  title = {Probing {{Real-Space}} and {{Time-Resolved Correlation Functions}} with {{Many-Body Ramsey Interferometry}}},
  author = {Knap, Michael and Kantian, Adrian and Giamarchi, Thierry and Bloch, Immanuel and Lukin, Mikhail D. and Demler, Eugene},
  year = 2013,
  month = oct,
  journal = {Physical Review Letters},
  volume = {111},
  number = {14},
  pages = {147205},
  publisher = {American Physical Society},
  doi = {10.1103/PhysRevLett.111.147205},
  url = {https://link.aps.org/doi/10.1103/PhysRevLett.111.147205},
  urldate = {2026-03-03}
}

@preamble{ "\providecommand{\noopsort}[1]{} " }
\let\addcontentsline\oldaddcontentsline

\clearpage
\widetext
\begin{center}
\textbf{\large Supplementary Materials:\\ \mytitle}
\end{center}

\tableofcontents
\setcounter{equation}{0}
\setcounter{figure}{0}
\renewcommand{\figurename}{FIG.}
\renewcommand{\thefigure}{S\arabic{figure}}
\renewcommand{\theequation}{S\arabic{equation}}
\renewcommand{\theHfigure}{S\arabic{figure}}
\makeatother

\section{Single-qubit evolution with arbitrary state preparation and readout in the presence of decoherence}

\begin{figure}
{\includegraphics[scale=1.25]{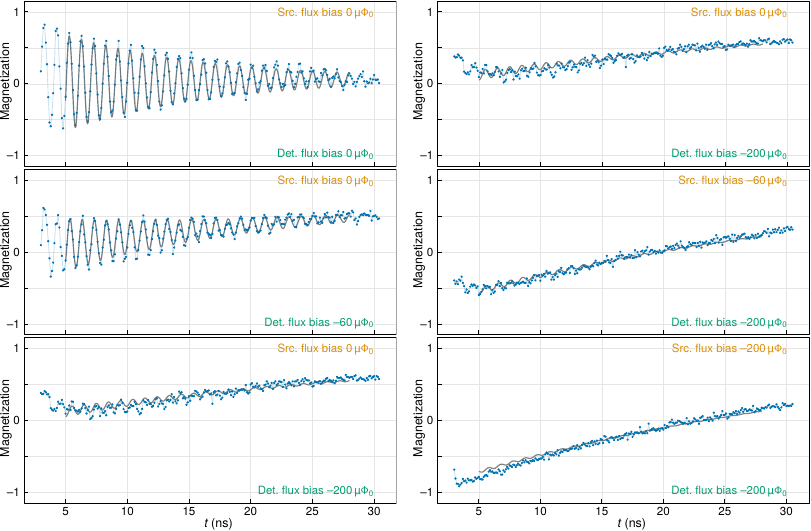}}
\caption{{\bf Single-qubit Larmor precession and fit.}  The six panels show the QPU data presented in Fig.~\ref{fig:2}b (blue dots) with pale blue lines as a guide to the eye.  Best-fit model~\eqref{eq:bloch_angles} using the qubit-median $T_1$ and $T_2$ values is plotted in gray, over the fitting region $\SI{5}{ns}\leq t\leq \SI{28}{ns}$.}
\label{fig:angle-fits}
\end{figure}

The density matrix of a single qubit can be expressed in the Bloch-vector form
\begin{align}
\label{eq_bloch_state}
\rho(t) = \frac{1}{2}(\mathbbm{1} + \vec{n}(t).\vec{\sigma}).
\end{align}

Consider a single qubit prepared in an arbitrary pure initial state $\vec{n}(0) = \hat{n}_s =  (\sin\theta_s\cos\varphi_s, \sin\theta_s\sin\varphi_s, \cos\theta_s)$ at time $t=0$. The qubit then evolves under the Hamiltonian $\mathcal{H}_0 = -\frac{\Delta}{2}\sigma^z$, in the presence of relaxation and pure dephasing. The evolution of this state is given by the Lindblad master equation
\begin{align}
\frac{d\rho}{dt} &= -\frac{i}{\hbar}[\mathcal{H}_0,\rho] + \sum_i \left(\mathcal{L}_i\rho \mathcal{L}_i^\dagger - \frac{1}{2}\{\mathcal{L}_i^\dagger\mathcal{L}_i,\rho\}\right),
\end{align}
where the Lindblad operators are $\mathcal{L}_0 = \sqrt{\frac{1}{T_1}}\sigma^-$ for relaxation to the ground state and $\mathcal{L}_1 = \sqrt{\frac{1}{2T_\varphi}}\sigma^z$ for pure dephasing. This equation can be solved analytically \cite{Lidar_notes_2019}, giving
\begin{align}
\label{eq_bloch_solution}
\vec{n}(t) = \begin{pmatrix}
[n_x(0) \cos(\omega t) - n_y(0) \sin(\omega t)]
e^{-\frac{t}{T_2}}
\\
[n_x(0) \sin(\omega t) + n_y(0) \cos(\omega t)]
e^{-\frac{t}{T_2}}
\\
1-[1-n_z(0)]e^{-\frac{t}{T_1}}
\end{pmatrix},
\end{align}
where $\omega = \Delta/\hbar$ and $\frac{1}{T_2} = \frac{1}{T_\varphi}+\frac{1}{2 T_1}$ is the total dephasing rate. Finally, the measurement along an arbitrary axis can be expressed by applying the rotation $R^{-1}_{\theta_d,\varphi_d}$ and then measuring in the computational basis, where $\hat{n}_d = (\sin\theta_d\cos\varphi_d,\sin\theta_d\sin\varphi_d,\cos\theta_d)$ is the unit vector specifying the measurement axis on the Bloch sphere. Equivalently, $\sigma_{\hat{n}_d} = R_{\theta_d,\varphi_d} \sigma_z R^{-1}_{\theta_d,\varphi_d}$, and the expectation value is $\Tr[\rho(t)\sigma_{\hat{n}_d}]$. After substituting $\rho(t)$ from Eq. \eqref{eq_bloch_state} and Eq.~\eqref{eq_bloch_solution} and some algebra, we arrive at Eq.~\eqref{eq:bloch_angles}. Fits to this model for the data in Fig.~\ref{fig:2}b are shown in Fig.~\ref{fig:angle-fits}.

\section{Two-qubit evolution under XY Hamiltonian and decoherence}

For a two-qubit system with the Hamiltonian in Eq.~\eqref{eq:H_2Q}, we can write the Lindblad master equation, assuming that there are two independent relaxation and dephasing processes on the two qubits but with the same timescales $T_1$ and $T_\varphi$. The Lindblad operators are then $\mathcal{L}_0 = \sqrt{\frac{1}{T_1}}\sigma^-_1$, $\mathcal{L}_1 = \sqrt{\frac{1}{2T_\varphi}}\sigma^z_1$, $\mathcal{L}_2 = \sqrt{\frac{1}{T_1}}\sigma^-_2$, $\mathcal{L}_3 = \sqrt{\frac{1}{2T_\varphi}}\sigma^z_2$.
This equation can be solved analytically, giving the exact solutions for an arbitrary initial state and measurement basis. For simplicity, we only consider the initial states prepared in the experiments in Fig.~\ref{fig:3} here. For $\psi(t=0) = \ket{10}$, we get
\begin{align}
\braket{\sigma_1^z} = 1 - e^{-\frac{t}{T_1}} 
- e^{-t (\frac{1}{T_1} + \frac{1}{T_\phi})} 
\left( 
\cosh (\frac{\nu t}{T_\phi}) + 
\frac{1}{\nu}\sinh(\frac{\nu t}{T_\phi}) 
\right),
\end{align}
where $\nu=\sqrt{1- \mathcal{J}^2T_\phi^2/\hbar^2}$. This expression can be simplified in the experimentally relevant regime $\mathcal{J}T_\varphi/\hbar \gg 1$, which is reasonable given that the estimated single-qubit value gives $\mathcal{J}T_\varphi/\hbar\ge 12$. With this approximation, we get
\begin{align}
\label{eq:sigmaz1}
\braket{\sigma_1^z} = 
1-e^{-\frac{t}{T_1}}(1+e^{-\frac{t}{T_\varphi}} \cos{\left(\mathcal{J} t/\hbar\right)}).
\end{align}
Similarly, we can write the other observables as
\begin{align}\label{eq:sigmaz2}
\braket{\sigma_2^z} &= 1-e^{-\frac{t}{T_1}}(1-e^{-\frac{t}{T_\varphi}} \cos{\left( \mathcal{J} t/\hbar\right)}),
\\
\braket{\sigma_1^z\sigma_2^z} &= 1-2e^{-\frac{t}{T_1}}.\label{eq:sigmazsigmaz}
\end{align}

\begin{figure}[t!]
{\includegraphics[scale=1.3]{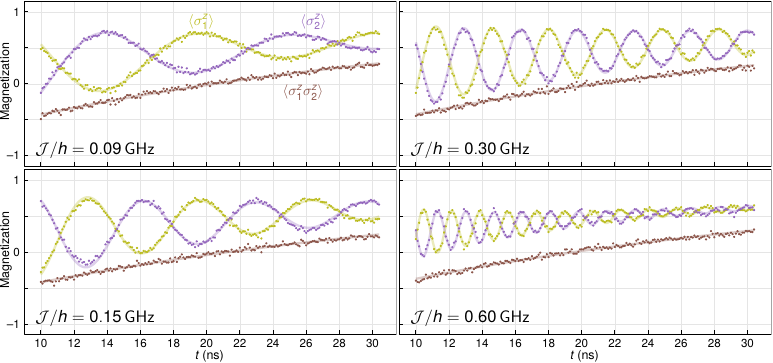}}
\caption{{\bf Two-qubit $\sigma^z$-basis fits with varying coupling.}  Data as in Fig.~\ref{fig:3} shows $\langle\sigma^z\rangle$ and $\langle\sigma^z\sigma^z\rangle$ expectations following a $\pi$ pulse for a single two-target-qubit system at $\Delta/h=\SI{1.0}{GHz}$, with $\mathcal J/h=\SI{0.15}{GHz}$ and $\mathcal J/h=\SI{0.30}{GHz}$ corresponding to the rightmost two columns of Fig.~\ref{fig:3}.  Experimental data are shown as small circles, each corresponding to the average of two 1000-shot QPU calls with opposite polarizing bias on the source qubit; fits to theory~Eqs.~\eqref{eq:sigmaz1}--\eqref{eq:sigmazsigmaz} are shown as solid lines.}
\label{fig:2q-fits}
\end{figure}

Fig.~\ref{fig:2q-fits} shows that the experimental data are in excellent agreement with this model, and gives $T_\varphi = \SI{37}{ns}$, resulting in a single-qubit coherence time $T_2 = \SI{23}{ns}$.

\section{Multi-level simulations of coupled rf-SQUIDs}

This section presents simulations of the Hamiltonians for inductively coupled, superconducting, compound-compound-Josephson-junction rf-SQUIDs (CCJJ-RFSs) to demonstrate the role of higher excited states in the CCJJ-RFSs during the excitation and measurement processes described in the main text. The QA QPU used in this study is composed of a network of CCJJ-RFSs that serve as qubits and compound-Josephson-junction rf-SQUIDs (CJJ-RFSs) that serve as inter-qubit couplers ~\cite{CCJJqubit,CJJcoupler,harris_experimental_2010}. The CJJ-RFSs comprising the couplers are designed to behave as classical objects that remain in their ground states at all times, with very large spectral gaps between ground and first excited states throughout their range of operating parameters. The CCJJ-RFSs comprising the qubits must necessarily provide two-level low-energy manifolds.  As a side effect, the higher excited states of these CCJJ-RFSs also occur at lower energies than the CJJ-RFS couplers. When the qubit CCJJ-RFSs are coupled via nominally strong mutual inductances, the low energy manifold of the resultant coupled system Hamiltonian matches that of the desired coupled qubit system to high accuracy. However, weak hybridization with the aforementioned higher excited states can lead to resolvable quantitative deviations.

It is demonstrated in \cite{CCJJqubit} that the 4-dimensional Hamiltonian of an isolated CCJJ-RFS can be reduced to a 2-dimensional Hamiltonian of the form
\begin{subequations}
\begin{equation}
\label{eq:HCCJJRFS}
{\cal H}_{\rm CCJJ-RFS}=\sum_{i\in\left\{\rm q,ccjj\right\}}\left[\frac{\hat{Q}^2_i}{2C_i}+\frac{\Phi^2_0}{4\pi^2L_i}\frac{\left(\hat{\varphi}_i-\varphi^x_i\right)^2}{2}\right]-\frac{I_c(\Phi^x_L,\Phi^x_R)\Phi_0}{2\pi}\cos\left(\frac{\hat{\varphi}_{\rm ccjj}-\varphi^0_{\rm ccjj}}{2}\right)\cos\left(\hat{\varphi}_q-\varphi^0_q\right),
\end{equation}
\begin{equation}
    \label{eq:Ic}
    I_c(\Phi^x_L,\Phi^x_R)=(I_{c1}+I_{c2})\cos\left(\frac{\pi\Phi^x_L}{\Phi_0}\right)+(I_{c3}+I_{c4})\cos\left(\frac{\pi\Phi^x_R}{\Phi_0}\right),
\end{equation}
\end{subequations}
where $\hat{\varphi}_i$ and $\hat{Q}_i$ are the phase and charge degrees of freedom for mode $i$ that obey the uncertainty relation $\left[\Phi_0\hat{\varphi}_i/2\pi,\hat{Q}_i\right]=i\hbar$ and $\left\{I_{cj}|1\leq j\leq4\right\}$ are the critical currents of the four JJs in the CCJJ-RFS.  For an appropriate choice of externally controlled flux biases $\Phi^x_L$ and $\Phi^x_R$, the modes $i=~$q and $i=~$ccjj can be associated with closed superconducting loops characterized by inductances $L_{\rm q}$ and $L_{\rm ccjj}$, respectively, and shunted by capacitances $C_{\rm q}$ and $C_{\rm ccjj}$, respectively.  These closed superconducting loops are subjected to externally controlled fluxes $\Phi^x_i$ that appear in Hamiltonian \ref{eq:HCCJJRFS} through the phase offsets $\varphi^x_i\equiv2\pi\Phi^x_i/\Phi_0$.  Junction asymmetry between members of pairs $\left\{I_{c1},I_{c2}\right\}$ and $\left\{I_{c3},I_{c4}\right\}$ results in additional phase offsets $\varphi^0_{\rm q}$ and $\varphi^0_{\rm ccjj}$ that appear in Hamiltonian \ref{eq:HCCJJRFS}.  In the limit $L_{\rm ccjj}\ll L_{\rm q}$ and $C_{\rm ccjj}\ll C_{\rm q}$ it is reasonable to approximate $\hat{\varphi}_{\rm ccjj}\approx\varphi^x_{\rm ccjj}$ and to ignore fluctuations in $\hat{Q}_{\rm ccjj}$.  With these approximations, Hamiltonian \ref{eq:HCCJJRFS} can be reduced to
\begin{equation}
    \label{eq:Hfq}
    {\cal H}_{RFS}\approx 4E_c\hat{n}^2_{\rm q}+\frac{1}{2}E_L\left(\hat{\varphi}_{\rm q}-\varphi^x_{\rm q}-\varphi^0_{\rm q}\right)^2-E_J(\Phi^x_{\rm ccjj})\cos\left(\hat{\varphi}_{\rm q}\right) ;
\end{equation}
where $\hat{n}_{\rm q}\equiv\hat{Q}_{\rm q}/2e$, $E_c\equiv e^2/(2C_{\rm q})$, $E_L\equiv\Phi^2_0/4\pi^2L_{\rm q}$, and $E_J(\Phi^x_{\rm ccjj})$ is a CCJJ bias-dependent Josephson energy.  This latter Hamiltonian, modulo the additional phase offset $\varphi^0_{\rm q}$ and the bias-dependent $E_J$, is a familiar form that is used to describe a wide variety of flux qubits.  However, we have found it important to retain the CCJJ degree of freedom to accurately model the physics of the CCJJ-RFSs that are in our QA QPUs.  As such, all calculations discussed within this appendix have been based upon the 2-dimensional Hamiltonian \ref{eq:HCCJJRFS}.  The CCJJ-RFS parameters were taken to be equal to the mean values obtained during QA QPU calibration: $L_{\rm q}=113\,$pH, $C_{\rm q}=166\,$fF, $L_{\rm ccjj}=14\,$pH, $C_{\rm ccjj}=10\,$fF, and $I_c(\Phi^x_L,\Phi^x_R)=4.94\,\mu$A.  The asymmetry-dependent phases $\varphi^0_{\rm q}$ and $\varphi^0_{\rm ccjj}$ are nulled during QPU calibration by appropriately redefining $\varphi^x_{\rm q}$ and $\varphi^x_{\rm ccjj}$, respectively, and can therefore be set to zero.
 
The key bias for controlling the annealing parameter $s$, as used on the horizontal axis of the annealing schedule plot in Fig.~\ref{fig:1}a, is the flux bias applied to the CCJJ loops of the qubit CCJJ-RFSs.  Let this flux applied to device $\alpha$ be denoted by $\Phi^x_{\rm ccjj}\to\Phi^{x,\alpha}_{\rm ccjj}$, where $\alpha$ will be replaced by `src', `det', or `target' to indicate a source, detector, or target device, respectively.  For the QPU used in this study, $\Phi^{x,\alpha}_{\rm ccjj}(s=0)=0.703\Phi_0$ and $\Phi^{x,\alpha}_{\rm ccjj}(s=1)=0.777\Phi_0$, and the mapping from intermediate $0\leq s\leq1$ to $\Phi^{x,\alpha}_{\rm ccjj}(s)$ is monotonic and weakly nonlinear.  It is important to recognize that the transverse field energy scale denoted by $A(s)$ in Fig.~\ref{fig:1}a becomes exponentially small for $s\gtrsim0.6$.  Furthermore, the qubits' coherence times drop dramatically in this regime as they become increasingly coupled to environmental flux noise.  Consequently, to simulate interesting coherent dynamics, it suffices to restrict numerical simulations to within $0\lesssim s \lesssim 0.7$, which roughly corresponds to $0.7\leq\Phi^{x,\alpha}_{\rm ccjj}/\Phi_0\leq0.75$.

In addition to the aforementioned CCJJ bias, each qubit CCJJ-RFS also possesses a body loop that can be flux biased by a quantity that will be denoted $\Phi^x_{\rm q}\to\Phi^{x,\alpha}_{\rm q}$.  This control can be used to deterministically prepare CCJJ-RFSs in localized flux states or to perform an adiabatic energy-basis-to-flux-basis mapping. This will be described in sections \ref{subsec:excitation} and \ref{subsec:detection}. 

The analysis in these sections involves identification of states of a system of two coupled flux qubits. Table \ref{tab:states} lists the definitions of single-qubit states that have been used herein. Two-qubit states will be written in the form $\ket{A_qB_{\rm target}}$ where $A,B\in\left\{0,1,L,R\right\}$ and $q\in\left\{\rm{src},\rm{det}\right\}$.

\begin{table}[ht]
  \begin{tabular}{|c||c||c||l|} \hline
    \bf{Symbol} & \bf{Equivalent} & \bf{Basis} & \bf{Description} \\ \hline\hline
    $\ket{0}$ & $=\frac{1}{\sqrt{2}}\left[\ket{L}+\ket{R}\right]$ & Energy & Qubit ground state with body biased to {\bf degeneracy} \\ \hline
    $\ket{1}$ & $=\frac{1}{\sqrt{2}}\left[\ket{L}-\ket{R}\right]$ & Energy & Qubit excited state with body biased to {\bf degeneracy} \\ \hline
    $\ket{L}$ & $=\ket{+}=\frac{1}{\sqrt{2}}\left[\ket{0}+\ket{1}\right]$ & Flux & Left-circulating persistent current state \\ \hline
    $\ket{R}$ & $=\ket{-}=\frac{1}{\sqrt{2}}\left[\ket{0}-\ket{1}\right]$ & Flux & Right-circulating persistent current state \\ \hline
  \end{tabular}
  \caption{\label{tab:states} Single flux-qubit state definitions.}
\end{table}

\subsection{\label{subsec:excitation} Excitation of a target using a source}

\begin{figure}
\includegraphics[scale=1]{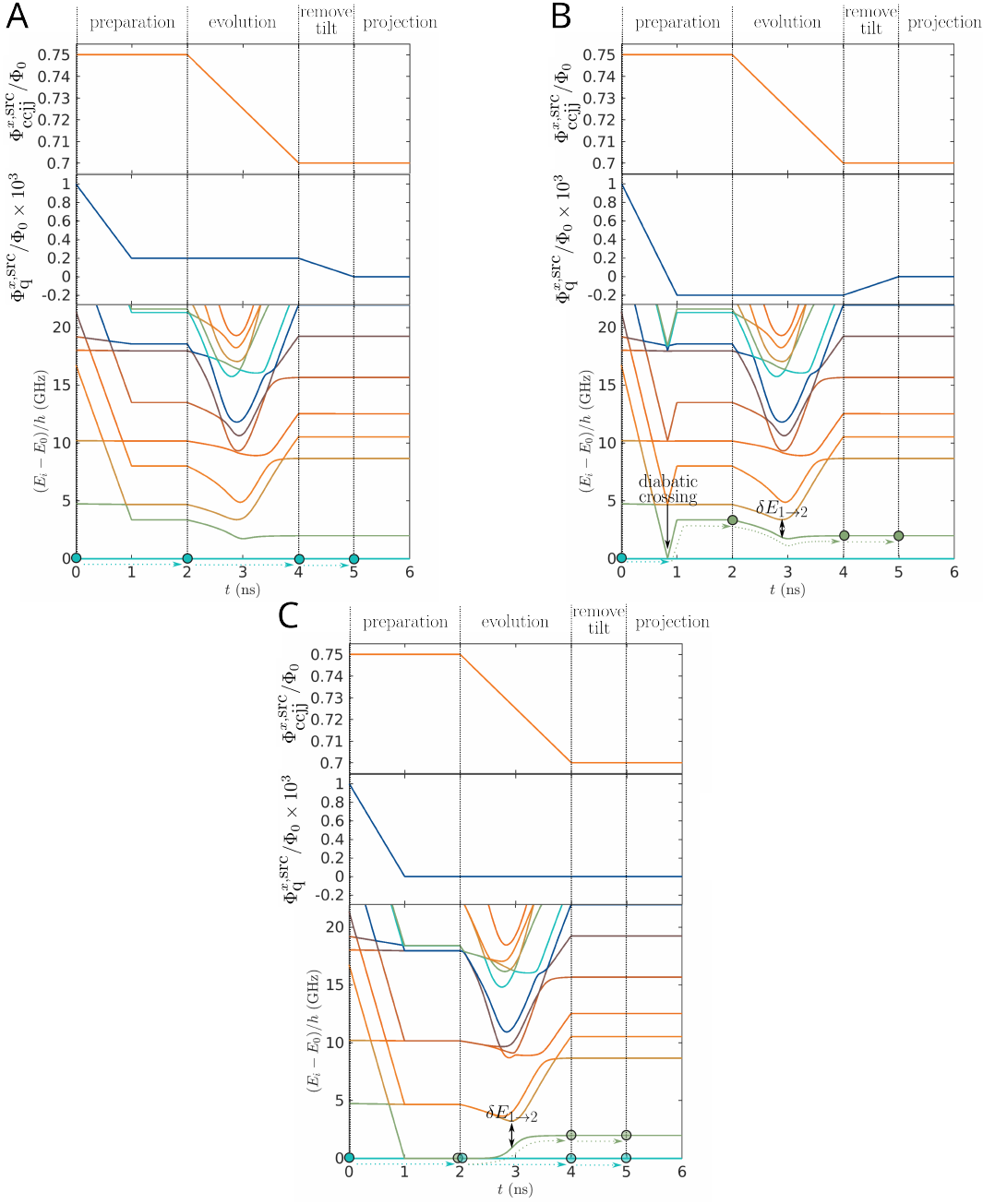}
\caption{\label{fig:source-eigenspectrum} Example control signals and instantaneous eigenspectra versus time for simulating a source CCJJ-RFS coupled to a target CCJJ-RFS.  The target is held at a constant CJJ bias and its body loop is biased at degeneracy throughout the process.}
\end{figure}

The excitation process can be modeled by considering a source CCJJ-RFS that is inductively coupled to a target CCJJ-RFS.  To simplify the analysis, the detector CCJJ-RFS is ignored. This simplification can be justified by noting that in practice the detector CCJJ-RFS is biased at $\Phi^{x,\rm det}_{\rm ccjj}=\Phi_0/2$ during target excitation. Thus, the detector's spectral gap is very large, i.e., on the order of $40\,$GHz. Under these conditions, the only effect of retaining the detector-target coupling is a very small inductive loading of the target. Assuming the CCJJ-RFS coupler can be adequately modeled by a lumped element mutual inductance, the required simulation involves only two CCJJ-RFSs representing the source and the target.  The instantaneous Hamiltonian for this coupled system can be written as
\begin{equation}
    \label{eq:HST}
    {\cal H}_{\rm src-target}(t)={\cal H}_{\rm CCJJ-RFS}\Bigl(\Phi^{x,\rm src}_{\rm ccjj}(t),\Phi^{x,\rm src}_{\rm q}(t)\Bigr)+{\cal H}_{\rm CCJJ-RFS}\left(\Phi^{x,\rm target}_{\rm ccjj},\Phi^{x,\rm target}_{\rm q}\right)+M\hat{I}^{\rm src}_p(t)\hat{I}^{\rm target}_p,
\end{equation}
where the src and target CCJJ-RFS Hamiltonians use identical device parameters as cited above, but src is subjected to time-dependent CCJJ and body flux biases and target is held at time-independent CCJJ and body flux biases.  The coupling is modelled using a lumped element mutual inductance $M=0.443\,$pH and the product of the persistent current operators for the CCJJ-RFSs
\begin{equation}
    \label{eq:Iqp}
    I^{\alpha}_p(t)\equiv\frac{\Phi_0}{2\pi L_q}\Bigl(\hat{\varphi}^{\alpha}_{\rm q}-\varphi^{x,\alpha}_{\rm q}(t)\Bigr) ,
\end{equation}
where $\hat{\varphi}^{\alpha}_{\rm q}$ is the q-mode phase operator and $\varphi^{x,\alpha}_{\rm q}(t)\equiv2\pi\Phi^{x,\alpha}_{\rm q}(t)/\Phi_0$ is the externally controlled q-mode phase offset for CCJJ-RFS $\alpha\in\left\{\rm src,target\right\}$.

Figure \ref{fig:source-eigenspectrum} shows an example of source control signals and instantaneous eigenspectra versus time for three different source-body-bias settings corresponding to three different target-state preparations. The excitation process is divided into 4 phases, denoted as preparation, evolution, remove-tilt, and projection.

Preparation begins at $t=0$ with a large source tunnel barrier, obtained by setting $\Phi^{x,\rm src}_{\rm ccjj}(t=0)=0.75\Phi_0$, and a modestly large nonzero body bias $\Phi^{x,\rm src}_{\rm q}(t=0)=1\,$m$\Phi_0$. The target qubit is subject to static biases $\Phi^{x,\rm target}_{\rm ccjj}=0.726\Phi_0$ and $\Phi^{x,\rm target}_{\rm q}=0$ (note the $\pi$-shifted operation of the CCJJ-RFS flux qubit) such that its qubit spectral gap is $\Delta_q/h=2.0\,$GHz when the device is considered in isolation. These initial conditions result in a coupled system ground state that is hard-polarized in an antiferromagnetically aligned flux state $\ket{L_{\rm src}R_{\rm target}}$. The tilt bias $\Phi^{x,\rm src}_{\rm q}$ is then linearly ramped to a new level by $t=1\,$ns.  In Fig.~\ref{fig:source-eigenspectrum}A, that new value is $\Phi^{x,\rm src}_{\rm q}(t=1\,\mbox{ns})=0.2\,$m$\Phi_0$ and possesses the same sign as the initial value at $t=0$.  As such, the initial state $\ket{L_{\rm src}R_{\rm target}}$ still corresponds to the ground state.  In Fig.~\ref{fig:source-eigenspectrum}B, that new value is $\Phi^{x,\rm src}_{\rm q}(t=1\,\mbox{ns})=-0.2\,$m$\Phi_0$ and possesses the opposing sign compared to the initial bias at $t=0$.  This flip in polarity causes the initial state $\ket{L_{\rm src}R_{\rm target}}$ to become the first excited state of the coupled system.  In Fig.~\ref{fig:source-eigenspectrum}C, the new value is $\Phi^{x,\rm src}_{\rm q}(t=1\,\mbox{ns})=0$ and the ground state becomes doubly degenerate. The initial state $\ket{L_{\rm src}R_{\rm target}}$ then corresponds to a superposition of the two ground states.

The evolution phase beginning at $t=2\,$ns is defined by the reverse quench of $\Phi^{x,\rm src}_{\rm ccjj}(t)$ from $0.75\Phi_0$ to $0.7\Phi_0$ over $2\,$ns. This timescale is representative of the apparatus, based on modeling of the room-temperature electronics and signal propagation along the analog CJJ bias line to the QPU at the mixing chamber of the dilution refrigerator. A key feature to note in the instantaneous eigenspectra within the evolution phase is the minimum in the first excited state to second excited state gap denoted as $E_{1\to2}$ in Figs.~\ref{fig:source-eigenspectrum}B and C.

The remove-tilt and projection phases involve adiabatically moving the coupled system to degeneracy by $t=5\,$ns and then projecting onto the coupled system's energy eigenstates at $t=6\,$ns. We note that the tilt removal was used in these simulations out of convenience; it was not present in the experiments. All other biases remain constant through the remove-tilt and projection phases. The ground and first excited states for $t>5\,$ns correspond to $\ket{0_{\rm src}0_{\rm target}}$ and $\ket{0_{\rm src}1_{\rm target}}$, respectively. Thus, the low energy manifold is spanned by the separable states $\ket{0_{\rm src}}\otimes\left\{\ket{0_{\rm target}},\ket{1_{\rm target}}\right\}$ and the final state of the system can be projected onto $\ket{0_{\rm src}}\otimes\left[\cos(\theta/2)\ket{0_{\rm target}}+\sin(\theta/2)e^{i\varphi}\ket{1_{\rm target}}\right]$. From such a projection, one can then parameterize the final state of the target as a Bloch vector possessing polar angle $\theta$ and azimuthal angle $\varphi$. The fidelity of the projection onto the target qubit subspace $\cal F$ is defined as the magnitude of that vector. The timings of both tilt-removal and projection are arbitrary and only lead to an overall offset in the azimuthal phase $\varphi$ of the final state.

Simulations of the initialization procedure were performed using Hamiltonian \ref{eq:HST} in a time-dependent Schr\"{o}dinger-equation solver. At each time step of the simulation, the energy eigenstates of the coupled system had to be calculated.  These eigenstates were expressed as superpositions of product states formed from the instantaneous energy eigenstates of the individual CCJJ-RFSs.  To study the impact of higher energy levels of the CCJJ-RFSs, simulations were run using varying numbers of per-CCJJ-RFS energy eigenstates $n_{\rm eigs}$.  In the limit of very small $M$ (weak coupling), the low energy eigenstates of the coupled system can be well approximated by superpositions of the qubit states, $n_{\rm eigs}=2$.  For modest values of $M$ one must use $n_{\rm eigs}>2$ to construct accurate representations of the coupled system eigenstates.

\begin{figure*}
\includegraphics[scale=1]{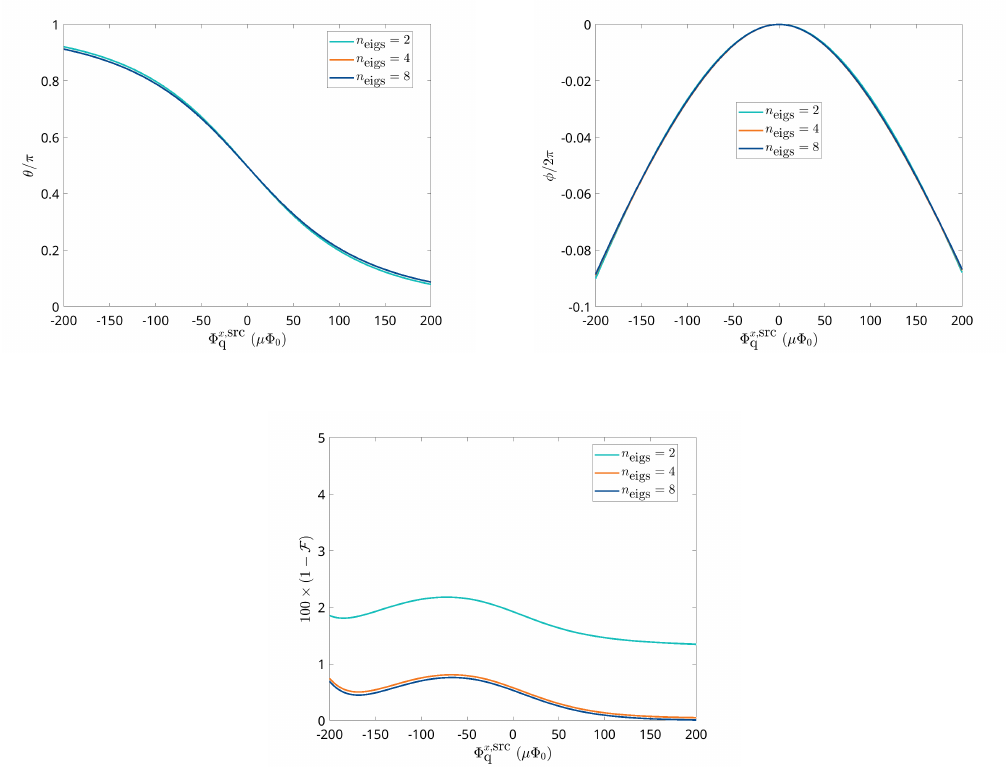}
\caption{\label{fig:source-simulation}  Projection of the final state of the excitation process in Bloch space as a function of source CCJJ-RFS body bias $\Phi^{\rm src}_{\rm q}$.  Results are shown for 3 values of the number of energy eigenstates per CCJJ-RFS $n_{\rm eigs}$  used to perform the coupled system simulation.}
\end{figure*}

Figure \ref{fig:source-simulation} summarizes the parameterization of the final state of the initialization procedure as a function of the flux bias applied to the source body $\Phi^{x,\rm src}_{\rm q}$ during the evolution phase.  Results are shown for three different values of $n_{\rm eigs}$.  The instantaneous eigenspectra shown in Fig.~\ref{fig:source-eigenspectrum} correspond to $n_{\rm eigs}=4$.  The $\theta$ and $\varphi$ results for $n_{\rm eigs}=4$ and $n_{\rm eigs}=8$ are coincident to within resolution whereas the $n_{\rm eigs}=2$ results exhibit small systematic differences.  Only the infidelity $1-{\cal F}$ exhibits an easily resolved disparity for $n_{\rm eigs}=2$.  Detailed investigation reveals that the aforementioned minimum spectral gap $E_{1\to2}$ is, perhaps surprisingly, smaller for $n_{\rm eigs}=2$ compared to $n_{\rm eigs}\geq4$.  The increased infidelity is due to the probability that the system experiences a diabatic transition out of the low energy manifold.  Either increasing the reverse quench time or decreasing the target qubit's spectral gap reduces the infidelity.  All of these observations are consistent with the higher excited states of the CCJJ-RFSs having a small but resolvable impact on the low energy eigenspectrum.

The results in Fig.~\ref{fig:source-simulation} indicate that each value of $\Phi^{x,\rm src}_{\rm q}$ results in a unique pair of angles $\left(\theta,\varphi\right)$.  However, note that simply adjusting the duration of the projection phase by $0\leq\delta t\leq h/\Delta$ facilitates arbitrary rotation of $\varphi$ by up to $2\pi$.  Thus, the excitation protocol presented herein can access the entire Bloch sphere.

The physics of the excitation protocol can be understood in greater detail by reviewing the instantaneous eigenspectra in Fig.~\ref{fig:source-eigenspectrum}. In all three cases shown, the system is initialized in the ground state with certainty, as indicated by the solid circle at $t=0$.  In Fig.~\ref{fig:source-eigenspectrum}A, wherein $\Phi^{x,\rm src}_{\rm q}>0$ in both the initialization and evolution phases, the system remains in the ground state and the final outcome is to leave the coupled system in the ground state.  Looking at the infidelity $1-{\cal F}$ in Fig.~\ref{fig:source-simulation}, one sees that this is trivially a high fidelity operation.  In contrast, the scenario depicted in Fig.~\ref{fig:source-eigenspectrum}B, wherein $\Phi^{x,\rm src}_{\rm q}$ changes sign, ends with the coupled system in its first excited state.  This action is facilitated by the diabatic crossing between the ground and first excited states at $t\approx0.85\,$ns.  Since $\Phi^{x,\rm src}_{\rm cjj}=0.75\Phi_0$ throughout the initialization phase, the spectral gap at this anticrossing is very small.  Consequently, the coupled system enters the evolution phase in the first excited state with certainty at $t=2\,$ns.  The only complication that can compromise the fidelity of this operation is the non-zero probability of making a second diabatic transition out of the low energy manifold across the minimum spectral gap $\delta E_{1\to2}$.  As can be seen in Fig.~\ref{fig:source-eigenspectrum}, there is a noticeable asymmetry in the infidelity $1-{\cal F}$ as a function of $\Phi^{x,\rm src}_{\rm q}$ that can be attributed to this additional diabatic excitation.  The most interesting case is illustrated in Fig.~\ref{fig:source-eigenspectrum}C wherein $\Phi^{x,\rm src}_{\rm q}=0$ for $t\geq1\,$ns.  The initial state entering the evolution phase can be expressed as a superposition of the two degenerate ground states $\ket{L_{\rm src}R_{\rm target}}=\left[\ket{+_{\rm src}+_{\rm target}}+\ket{+_{\rm src}-_{\rm target}}\right]/\sqrt{2}$, as indicated by the partially transparent circles at $t=2\,$ns.  The reverse quench lifts the degeneracy of the low energy manifold and the coupled system ends in a superposition state with equal amplitudes.  Diabatic transitions out of the low energy manifold are possible in the vicinity of the minimum spectral gap $\delta E_{1\to2}$, as shown in Fig.~\ref{fig:source-simulation}.

The preparation of an initial state possessing $\theta=0$ (no excitation) or $\theta=\pi$ (complete excitation or $\pi$ pulse) does not require phase coherence.  In both of these limits, the simple pictures based upon adiabaticity presented above will suffice.  However, the case for partial excitation $\theta=\pi/2$ ($\pi/2$ pulse) requires phase coherence.  The simulation results presented in Fig.~\ref{fig:source-simulation} were generated using a closed system model, so it is no surprise that the model yielded the observed behavior.  However, the physical system used in the experiments starts the evolution deep in the incoherent regime, where the states of the coupled system can be readily described in their flux bases, and makes a crossover to the coherent regime by the end of the reverse quench of the source.  The fact that the excitation process works in practice indicates that the details of the aforementioned crossover appear to be of little relevance.  In effect, the initial flux state $\ket{L_{\rm src}R_{\rm target}}$ is instantaneously projected onto the superposition states $\ket{+_{\rm src}+_{\rm target}}$ and $\ket{+_{\rm src}-_{\rm target}}$, provided the reverse quench is sufficiently fast.  The authors of \cite{Poletto_2009} reached similar conclusions using a related microwave-free method for applying fast control pulses to flux qubits.

\subsection{\label{subsec:detection} Reading a target using a detector}

\begin{figure}
\includegraphics[scale=1]{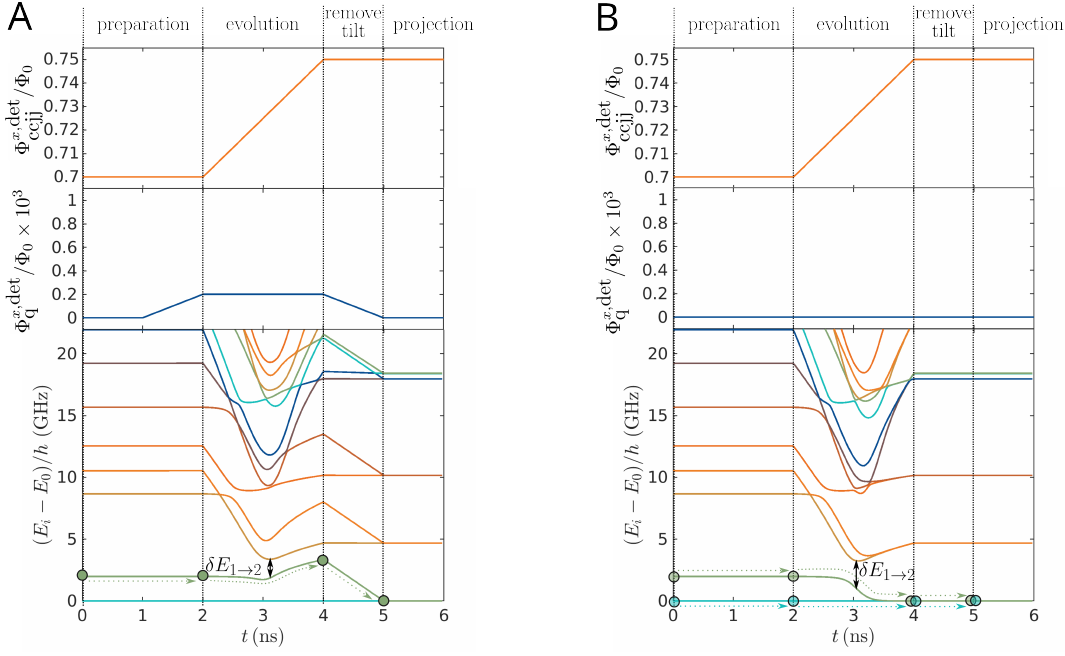}
\caption{\label{fig:detector-eigenspectrum} Example control signals and instantaneous eigenspectra versus time for simulating a target CCJJ-RFS coupled to a detector CCJJ-RFS.  In {\bf A}, the detector is biased to read the target in its energy basis.  In {\bf B}, the detector is biased to read the target in its flux basis.}
\end{figure}

To simulate readout using a detector CCJJ-RFS, one can neglect the source CCJJ-RFS because the latter is biased at $\Phi^{x,\rm src}_{\rm ccjj}=\Phi_0/2$ during the readout process.  The coupled system Hamiltonian can be written as
\begin{equation}
    \label{eq:HTD}
    {\cal H}_{\rm target-det}(t)={\cal H}_{\rm CCJJ-RFS}\left(\Phi^{x,\rm target}_{\rm ccjj},\Phi^{x,\rm target}_{\rm q}\right)+{\cal H}_{\rm CCJJ-RFS}\Bigl(\Phi^{x,\rm det}_{\rm ccjj}(t),\Phi^{x,\rm det}_{\rm q}(t)\Bigr)+M\hat{I}^{\rm target}_p\hat{I}^{\rm det}_p(t),
\end{equation}
with the CCJJ-RFS device parameters cited previously, $M=0.443\,$pH, and the persistent current operator $I^{\alpha}_p(t)$ for device $\alpha$ defined in Eq.~\ref{eq:Iqp}.

Figure \ref{fig:detector-eigenspectrum} shows example control signals and instantaneous eigenspectra for the limiting cases of measuring the target in its energy and flux bases.  In both cases, the system starts at $t=0$ with the detector biased at $\Phi^{x,\rm det}_{\rm ccjj}=0.7\Phi_0$.  The target qubit is subject to static biases $\Phi^{x,\rm target}_{\rm ccjj}=0.726\Phi_0$ and $\Phi^{x,\rm target}_{\rm q}=0,$ such that its spectral gap is $\Delta/h=2.0\,$GHz when the device is considered in isolation.  During the evolution phase, the detector CJJ bias $\Phi^{x,\rm det}_{\rm ccjj}$ is ramped from $0.7\Phi_0$ to $0.75\Phi_0$ within $2\,$ns.  To measure in the energy bases, the detector is pulsed to nonzero $\Phi^{x,\rm det}_{\rm q}$ during the evolution phase.  To measure in the flux bases, the detector's body bias is held at $\Phi^{x,\rm det}_{\rm q}=0$.  The difference in detector body bias leads to significantly different low energy eigenspectra in the evolution phase: for the energy basis measurement, the energy splitting between ground and first excited state remains large; for the flux basis measurement, the low energy manifold becomes a doubly degenerate ground state.  $\Phi^{x,\rm det}_{\rm q}$ finally returns to zero by $t=5\,$ns in all cases.  Finally, the protocol finishes at $t=6\,$ns with a calculation of the probability of observing the detector in the states $\ket{L_{\rm det}}$ and $\ket{R_{\rm det}}$.

\begin{figure*}
\includegraphics[scale=1]{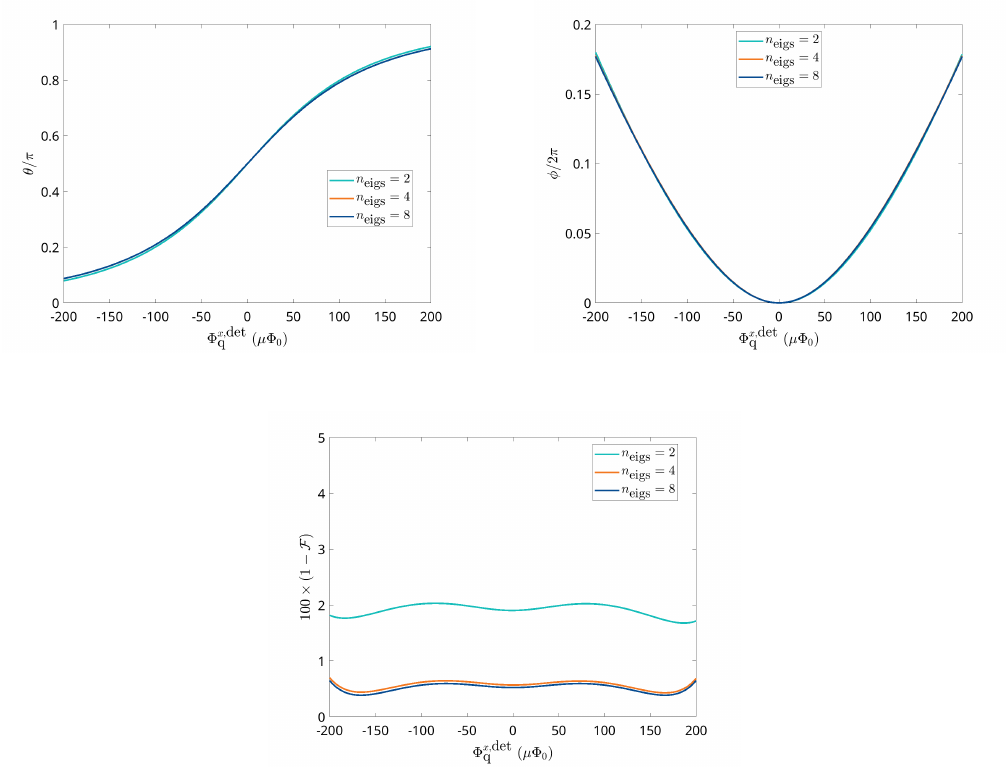}
\caption{\label{fig:detector-simulation} Simulation of target-detector protocol as a function of flux bias applied to the body of the detector rf-SQUID.  Results are shown for 3 different numbers of eigenstates per CCJJ-RFS.}
\end{figure*}

The readout process was characterized as a function of $\Phi^{x,\rm det}_{\rm q}$ pulse amplitude in the evolution phase for multiple values of $n_{\rm eigs}$.  Quantum process tomography was performed for each set of simulation conditions.  The tomography results were then summarized in terms of an axis in Bloch space along which measurement of the target had been performed, as characterized by polar angle $\theta$, azimuthal angle $\varphi$, and fidelity $\cal F$.  The results are summarized in Fig.~\ref{fig:detector-simulation}.  The angle results for the readout simulations are nearly coincident in all cases and only infidelity $1-{\cal F}$ exhibits an easily discernible difference for $n_{\rm eigs}=2$.  As in the excitation process, the timing of the readout sequence can be adjusted to realize measurement along an arbitrary axis on the Bloch sphere. In particular, adjusting the duration of the preparation phase rotates $\varphi$, allowing access to arbitrary pairs $\left(\theta,\varphi\right)$.

The annotation in the eigenspectra panel in Fig.~\ref{fig:detector-eigenspectrum}A illustrates the evolution of the system after having been initialized in the first excited state.  The large spectral gap within the low energy manifold ensures that no probability is transferred to the ground state up to $t=4\,$ns.  Thereafter, all dynamics in the coupled system have been quenched, the tilt applied to the detector is removed, and the system finishes in the antiferromagnetically aligned flux basis state $\ket{L_{\rm det}R_{\rm target}}$ with high fidelity.  In this case, the readout process amounts to a $\delta\theta=\pi/2$ rotation from the energy bases to the flux bases, followed by projection into the flux bases of the coupled system.  This process bears some resemblance to historical implementations of the tilt-and-latch readout of flux qubits \cite{earlyfluxqubit}, although the tilt and latch actions are now combined into a single operation controlled by the detector quench.

The annotation in the eigenspectra panel in Fig.~\ref{fig:detector-eigenspectrum}B illustrates the evolution of the system after having been initialized in the superposition state $\ket{+_{\rm det}}\otimes\left[\ket{+_{\rm target}}+\ket{-_{\rm target}}\right]/\sqrt{2}$.  Provided the evolution remains adiabatic, that superposition will survive into the projection phase, albeit with an accrued azimuthal phase difference $\delta\varphi$ that depends on the details of that evolution.  The final superposition then projects onto the coupled system flux bases $\ket{L_{\rm det}R_{\rm target}}$ and $\ket{R_{\rm det}L_{\rm target}}$.  In this case, one achieves readout of the target in its flux bases.  Similar to the $\pi/2$ pulse initialization process, in practice, readout in the flux bases involves a crossover from coherent to incoherent dynamics.  Provided that the forward quench of the detector is sufficiently fast, it appears that the projection into the localized flux bases can be considered instantaneous.

\section{Non-idealities in detection using an auxiliary qubit}

In this section, we analyze measurement using an auxiliary qubit within a two-level model and examine the effect of leaving the coupling between computational qubits on during measurement. In the notation of Eq.~\eqref{eq:H_anneal}, the detector qubit is deterministically initialized in the ground state $\ket{+}$ at $s=0$. At the desired point in time where we wish to measure the target qubit state, we now anneal the detector to $s=1$ with a fast ramp and measure the final state of the detector at $s=1$ in the computational basis through a standard persistent current measurement. This process can be treated as an effective map on a general target qubit using the superoperator formalism. First, the preparation of the detector qubit in $\ket{+}$ corresponds to
\begin{align}
s_{\rm prep} &= \frac{1}{2}\sum_{i,j,k,l,p,q}\delta_{ip}\delta_{kq} \ket{ijkl}\bra{pq}
\nonumber
\\
&= \frac{1}{2}\sum_{i,j,k,l}\ket{ijkl}\bra{ik},
\end{align}
which is a $16 \times 4$ superoperator. Next, we can write the propagator for the coupled two-qubit system through the forward anneal of the detector as
\begin{align}
u_{\rm prop} &= \mathcal{T}
e^{-\frac{i}{\hbar} \int H (t) dt}
\\
s_{\rm prop} &= \mathrm{u^{\ast}_{\rm prop}}\otimes\mathrm{u_{\rm prop}},
\end{align}
by using Roth's column lemma, where $\mathcal{T}$ is the time-ordering operator. $s_{\rm prop}$ is a $16 \times 16$ superoperator that acts on two-qubit operators. Since we ultimately measure only the detector qubit, we need to take the partial trace over the target qubit at the end of the anneal. This can also be written as a superoperator that performs the tensor contraction, given by a $4\times 16$ matrix:
\begin{align}
    s_{\rm ptrace} &= \sum_{\{i,j,k,l,p,q\}}\delta_{ik}\delta_{pj}\delta_{ql} \ket{pq}\bra{ijkl}\\
    &= \sum_{\{i,j,l\}} \ket{jl}\bra{ijil}.
\end{align}

At the end of the anneal, the state of the detector qubit rapidly dephases into the computational basis. This corresponds to applying the $4 \times 4$ superoperator:
\begin{align}
    s_{\rm dephasing} = \sum_i \ket{ii}\bra{ii}.
\end{align}
We can now combine these superoperators to calculate the effective process for an arbitrary initial state on the target qubit mapped onto the computational basis of the detector at the end. For this calculation, we transform the final $4 \times 4$ superoperator
\begin{align}
s_{\rm tot} = s_{\rm dephasing}*
s_{\rm ptrace}
*s_{\rm prop}*
s_{\rm prep}
\end{align}
into the Pauli basis, and extract the vector $\{\ket{Z}\bra{X},\ket{Z}\bra{Y},\ket{Z}\bra{Z}\}$. This vector quantifies the axis of the target qubit that is mapped onto the detector $\sigma^z$-basis, and its norm provides readout fidelity. The results for a single target-detector system agree qualitatively with the rf-SQUID simulations.

\begin{figure}[t!]
\centering
\includegraphics[scale=0.32]{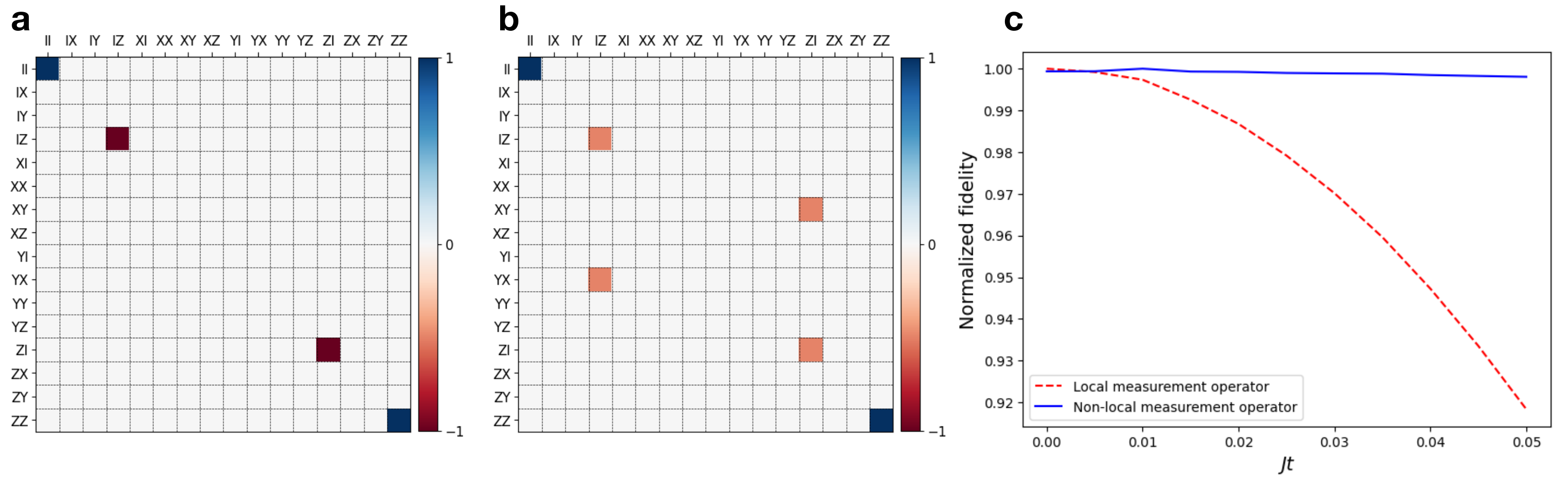}
\caption{{\bf Simultaneous two-qubit measurement superoperator.} {\bf a}, Qubits decoupled during measurement. {\bf b}, Finite target-target coupling during measurement. {\bf c}, Fidelity of local and non-local measurement operators.}
\label{fig:2q-detection-sim}
\end{figure}

In the current implementation of detector qubits using multicolor annealing, the coupling between target qubits is always on and follows the anneal schedule $B(s)$. In particular, the coupling is active during the detection process itself, which introduces a source of error in the effective local measurement operator realized by the protocol above. It is straightforward to extend the superoperator representation to two target qubits, each measured by their own detector, and calculate the $16 \times 16$ superoperator representing the process for a pair of target qubits with finite coupling. Now, separable components (e.g. $\{\ket{ZI}\bra{XI},\ket{ZI}\bra{YI},\ket{ZI}\bra{ZI}\}$) represent an effective ``local" measurement operator on the individual target qubits, and we can also calculate the fidelity of the two-qubit (``non-local") measurement operator. Fig. \ref{fig:2q-detection-sim} shows that while the fidelity of the local measurement operator drops as a function of the coupling strength between two target qubits, when we consider the non-local effective measurement operator realized by simultaneously measuring both qubits, we recover a significant portion of this fidelity. This behavior is in agreement with the intuition that the detector will be affected by neighboring qubit states when the timescale of detection is comparable to the coupling timescale to these qubits. So, as the timescales become comparable, the measurement performed by each detector qubit is no longer a local measurement of its specific target qubit but a measurement of the combined state of the two target qubits. We note that the non-local measurement still has finite infidelity, since the information is not mapped purely onto the  computational basis of the detectors, and dephasing of the detectors leads to some irreversible loss of information.

\section{QPU methods}\label{sec:qpumethods}

\begin{figure}
{\includegraphics[trim = 32mm 32mm 32mm 32mm, clip, width = 0.6\columnwidth]{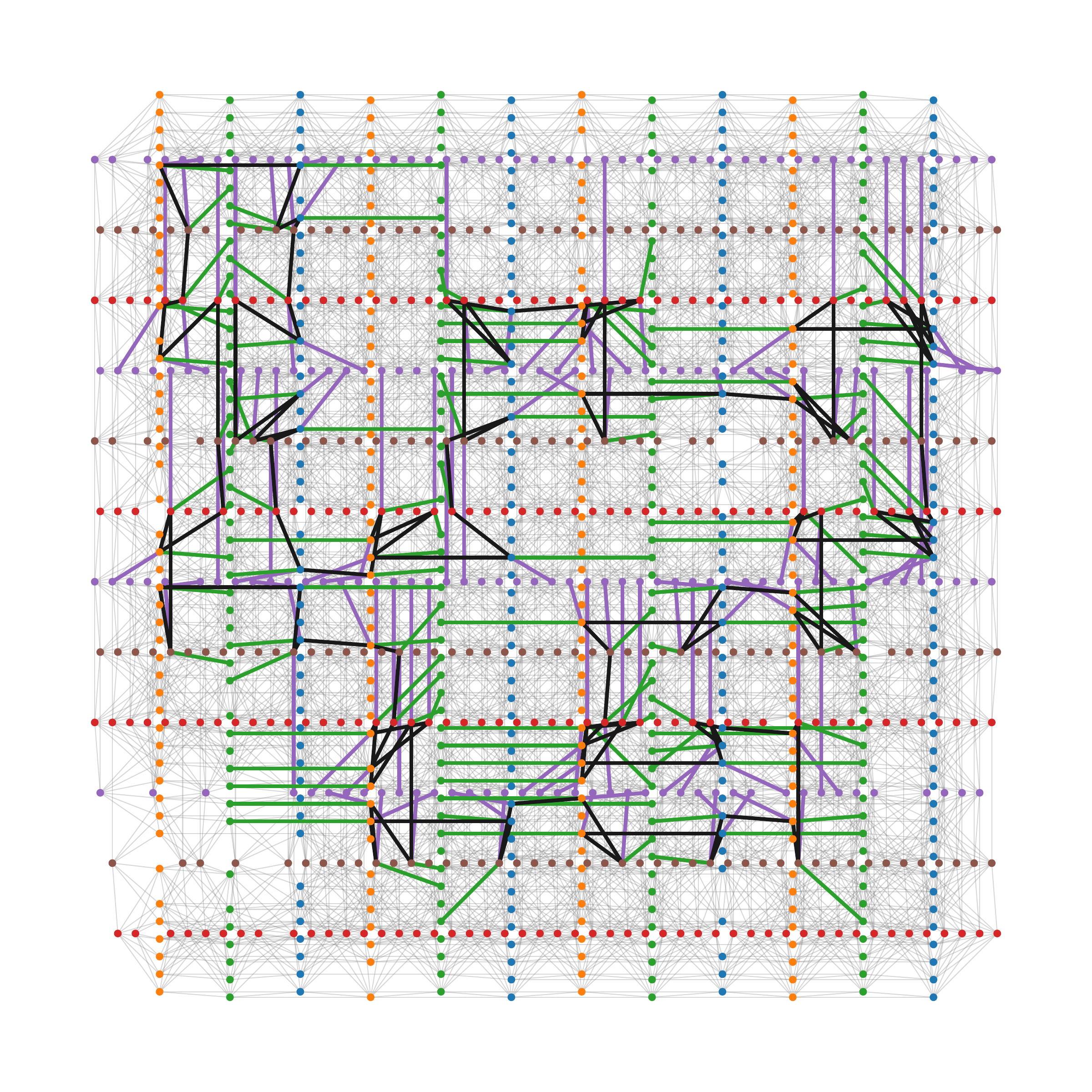}}
\caption{{\bf Qubit connectivity graph with six color classes.}  In multicolor annealing protocols, each color class can follow an independent anneal.  Shown is the qubit connectivity graph of the 1178-qubit Advantage2\texttrademark\ processor on which the experiments were performed, with a chain of length $L=124$ (bonds shown in black) embedded using four color classes for the target qubits (orange, red, brown and blue).  Each target qubit in the chain is coupled to a distinct detector qubit and source qubit from the two remaining color classes (purple and green respectively). Unused couplers are shown in light gray.}
\label{fig:processor-graph-colors}
\end{figure}

All QPU experiments shown in the main matter are performed through D-Wave's Python-based API using the Ocean\texttrademark\ Software Development Kit~\cite{dwavequantumOcean}.  Specific documentation on multicolor annealing experiments and a worked example on Larmor precession are available online~\cite{dwavequantumExperimentalResearch}.

All experiments are run on systems derived from periodic chains, either $L=124$ or $L=56$, consisting of $L$ target qubits, each associated with its own distinct source and detector qubit.  An embedding of such a system on 372 qubits is shown in Fig.~\ref{fig:processor-graph-colors}, with the six annealing lines indicated.

Calibration refinement of three-qubit source-target-detector systems is performed to synchronize oscillations and balance at a nominal degeneracy point (as in Fig.~\ref{fig:2}b) where we define zero values for per-line delays, per-qubit flux biases, and per-qubit anneal offsets.

In a first step, per-line baseline delays are determined so that the timing of the Larmor precession as a function of $t$ is roughly independent of which of the six annealing lines are used for target, source and detector.  This is done by running experiments for all three-tuples of lines and synchronizing their median time-series curves.

In a second step, per-qubit oscillation frequencies are homogenized using anneal offsets ($\delta s$) on target qubits.  Fine per-qubit synchronization of oscillation (following rough per-line synchronization) is achieved by tuning anneal offsets on source and target qubits; this uses a vertical shift in $s(t)$ to realize a horizontal shift in the intersection of the finite-slope source and detector quenches with the quantum-critical region of the target qubits.

In a third step, per-qubit flux biases are then used to balance the three-qubit systems at a degeneracy point corresponding to an unbiased target qubit initialized at $\theta_s=\pi/2$ and measured at $\theta_d=\pi/2$.

The second and third steps are repeated in several iterations. In a further calibration refinement step, measurements of $\mathcal J$ from coupled $\sigma^z$ oscillation frequencies as in Fig.~\ref{fig:3} could be used to homogenize couplings with high precision.

Data points in Fig.~\ref{fig:2}b and Fig.~\ref{fig:3} represent a single qubit's magnetization (or, in the case of $\langle\sigma^z\sigma^z\rangle$ in Fig.~\ref{fig:3}, the average of a two-qubit product) as measured from a 1000-shot QPU call for each value of $t$.  In Fig.~\ref{fig:4}, magnetizations in Fig.~\ref{fig:4}a and Fig.~\ref{fig:4}c are averaged over all possible locations of the initially excited qubit, and are averaged over the bracelet (site order reversal) symmetry of the chain.

\begin{figure*}[t!]
\includegraphics[scale=1.35]{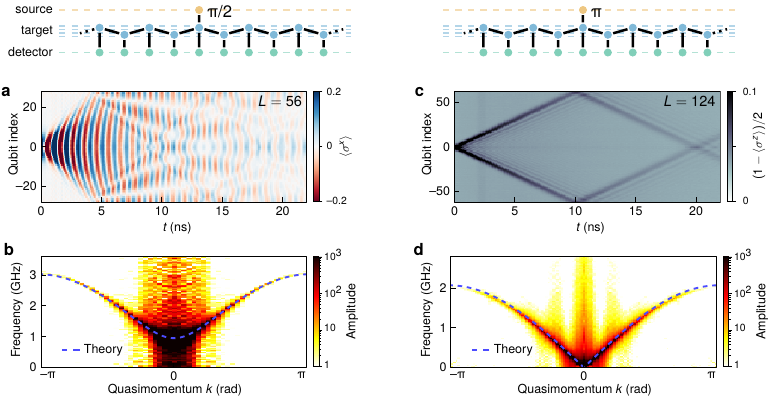}
\caption{{\bf QPU data for excitation propagation at strong coupling.} Data and experiment are as in Fig.~\ref{fig:4} with a stronger coupling $\mathcal J/h=\SI{-1.0}{GHz}$ instead of $\mathcal J/h=\SI{-0.6}{GHz}$.  {\bf a--b}, An excitation is prepared in the state $\ket +$ and measured in the $\sigma^x$ basis.  {\bf c--d}, An excitation is prepared in the state $\ket 1$ and measured in the $\sigma^z$ basis.}
\label{fig:propagation_1ghz}
\end{figure*}

Fig.~\ref{fig:propagation_1ghz} shows data analogous to Fig.~\ref{fig:4} at a stronger coupling $\mathcal J/h=\SI{-1.0}{GHz}$.  The greater background intensity in the $\sigma^z$-basis reflects the ground state of the chain.

When probing Anderson localization in Fig.~\ref{fig:5}, it is not possible to directly program per-qubit $\Delta$ detunings in~\eqref{eq:H_eff}.  Instead, we program per-qubit anneal offsets, which detune $s$, and therefore both $\Delta$ and $\mathcal J$.  We then modify programmed $\mathcal J_{ij}$ values to compensate for unwanted $\mathcal J$ detuning.  In the TFIM language of~\eqref{eq:H_anneal} and~\eqref{eq:HP_anneal}, we reprogram $J_{ij}$ values to compensate for the unwanted shifts in $B(s)$, as in Ref.~\cite{Lanting_2017_nonuniform_driver}.

\section{Theory of Magnetization Measurements of one-dimensional chains in the single-excitation approximation}
\label{sec:single-particle}

In the main text, we presented magnetization measurements as a function of time in two measurement bases ($x$ and $z$) for the clean transverse-field Ising chain (or equivalently, an XY Hamiltonian chain in the rotating wave approximation) initialized to have a single excitation (see Fig.~\ref{fig:4}).
These two propagation patterns (corresponding to the two measurement bases) have Fourier transforms that yield the dispersion relation and the energy-difference joint density of states.
Here, we derive these relationships within the single particle approximation, and provide theoretical predictions for the corresponding correlation functions and their Fourier transforms.

We will find it convenient to use the transverse-field Ising Hamiltonian of the form Eqs.~\eqref{eq_h0} and \eqref{eq_hi}, instead of the effective XY Hamiltonian in Eq.~\eqref{eq:H_eff} of the main text. The experiment therefore corresponds to the Hamiltonian
\begin{align}
    H = \frac{\mathcal J}{2}\sum_{i=0}^{L-1} \sigma_i^x \sigma_{i+1}^x  - \frac{\Delta }{2} \sum_{i=0}^{L-1} \sigma_i^z. \label{eq:TFIM_chain_appendix}
\end{align}
Here $i+1$ is understood modulo $L$, corresponding to periodic boundary conditions. In the weak coupling limit where $\frac{\mathcal{J}}{\Delta} \rightarrow 0$, the lowest energy sector is spanned by a single state $\ket 0_v \coloneqq \ket{0}^{\otimes L}$, which has all qubits aligned along the $z$ direction.
The first excited energy sector comprises all states with a single qubit being in state $\ket 1$, with each remaining qubit being in the state $\ket 0$.
We will denote the state with an excitation at location $n$ by $\ket{n}_e$.

Although it would be desirable to predict the magnetization measurements using the exact solution of Eq.~\eqref{eq:TFIM_chain_appendix}, the calculations are complicated by the non-locality of the $\sigma^x$ operator in terms of the Jordan-Wigner fermions.
However, a much simpler analysis, involving effective Hamiltonians for the two lowest energy sectors (to which the initial states of the two experiments belong) provides excellent agreement with the experimental results.

To first order in $\frac{\mathcal J}{\Delta}$, effective Hamiltonian $H_\text{eff}^{(1)}$ for the single excitation sector is given by~\cite{Sachdev_2011}
\begin{align}
    H_\text{eff}^{(1)} = \Delta \sum_n \ket{n}_e \bra{n} + \frac{\mathcal J}{2} \sum_{n=0}^{L-1} (\ket{n+1}_e \bra{n}_e + \ket{n}_e \bra{n+1}_e),
\end{align}
up to a constant that ensures that $H_\text{eff}^{(0)}=0$ for the ground state sector.
This Hamiltonian is easily diagonalized by defining magnon states $\ket{k}_e\coloneqq \frac{1}{\sqrt{L}} \sum_n e^{ikn} \ket{n}_e$, using which, we may write
\begin{align}
H_\text{eff}^{(1)} &= \sum_k E_\text{eff}^{(1)}(k) \ket{k}_e \bra{k}_e,
\end{align}
where
\begin{align}
E_\text{eff}^{(1)}(k) &= \Delta + \mathcal J \cos(k). 
\label{eq:E_eff}
\end{align}
As expected, the exact dispersion relation $E_\text{exact}(k)$ (obtained using the Jordan-Wigner transform) matches $E_\text{eff}^{(1)}(k)$ to the first order in $\frac{\mathcal J}{\Delta}$ as follows
\begin{align}
E_\text{exact}(k) = \sqrt{(\Delta + \mathcal J \cos k)^2 + \mathcal J^2 \sin^2 k} = E_\text{eff}^{(1)}(k) + \Delta \left[\mathcal O\left( \frac{\mathcal J^2}{\Delta ^2}\right) \right].
\end{align}
It is worth noting that the resulting group velocity is given by
\begin{equation}
v = \frac{1}{\hbar}
\pdv{}{k}
E_\text{eff}^{(1)}(k)
= 
-\frac{\mathcal{J}}{\hbar}
\sin (k)
\end{equation}
in the weak coupling limit. So, the maximum group velocity equals $\abs{\mathcal{J}}/ \hbar$ since we set the lattice constant to unity after the Jordan-Wigner transform. Consequently, $\mathcal{J}t/ \hbar$ is dimensionless, and defines the light-cone for the propagation of information in the system.

\subsection{\texorpdfstring{$\sigma^x$}{x basis} measurements}
With the initial state set to $\ket{\psi(t=0)}=\frac{1}{\sqrt{2}}(\ket{0}_v + \ket{0}_e) \equiv \frac{1}{\sqrt{2}}(1 + \sigma_0^x) \ket{0}_v$, we experimentally measured $m_x(n,t) = \bra{\psi(t)}\sigma_n^x \ket{\psi(t)} $ for all qubits $n$ over a fine grid of time values (see Figs.~\ref{fig:4}a,b in the main text). Equivalently, we may express this quantity as a zero-temperature retarded Green's function: $m_x(n,t) = iG_{xx}^{R+}(n,t) =  \frac{\theta(t)}{2} \bra{0}_v \{\sigma_n^x(t), \sigma_0^x  \}\ket{0}_v$.
Other Green's functions may be extracted via many-body Ramsey interferometry experiments, by changing the orientation of the initial excitation and of the measurement basis~\cite{KnapProbing2013}.
Here, we restrict to $G_{xx}^{R+}$.
(Let us note that $\ket{0}_v$ is an approximate ground state in the $\Delta \gg \mathcal J$ limit, to which we have restricted our discussion.)

Using the first order approximation, we obtain
\begin{align}
\ket{\psi(t)} &\approx \frac{1}{\sqrt{2}} \big(\ket{0}_v + e^{-iH_\text{eff}^{(1)}t/\hbar}\ket{0}_e
\big) 
\nonumber
\\&=
\frac{1}{\sqrt{2}} \Big(
\ket 0_v + \frac{1}{\sqrt{L}} \sum_k e^{-iE_\text{eff}^{(1)}(k)t/\hbar} \ket{k}_e \Big).
\end{align}
Restricting to the zero and one excitation sectors, we note that $\sigma_n^x \ket{0}_v = \ket{n}_e$, while $\bra{0}_v\sigma_n^x \ket{m}_e = \delta_{nm}$.
This immediately yields
\begin{align}
m_x(n,t) = \frac{1}{L} \sum_k 
\cos \Big(
E_\text{eff}^{(1)} (k)t/\hbar + kn
\Big).
\label{eq:m_x_single_photon}
\end{align}
A Fourier transform in space (discrete) and time (continuous)\footnote{We use the convention $\widetilde m(k,\omega) = \sum_n \int_{-\infty}^{\infty} e^{i(kn+\omega t)}\ m(n,t)\ \dd t$.} yields
\begin{align}
    \widetilde m_x(k,\omega) = \pi [\delta (\omega-E_\text{eff}^{(1)}(k)/\hbar) + \delta(\omega + E_\text{eff}^{(1)}(k)/\hbar)], \label{eq:mx_tilde_single_particle}
\end{align}
which implies that the peak should overlap with the following single-particle dispersion relation:
\begin{align}
    \omega_\text{peak}(k) = \pm E_\text{eff}^{(1)}(k)/\hbar. \label{eq:omega_sigma_z_single_particle}
\end{align}
While this expression is exact in the single-particle approximation for the weak coupling regime, we expect (also based in general from the spectral decomposition of the correlation function) peaks at $\pm E_\text{exact}(k)/\hbar$ instead away from the weak coupling regime, which we indeed observed in the quantum annealer data shown in the main text.

\subsection{\texorpdfstring{$\sigma^z$}{z basis} measurements}
\begin{figure*}[t!]
\includegraphics[scale=1.3]{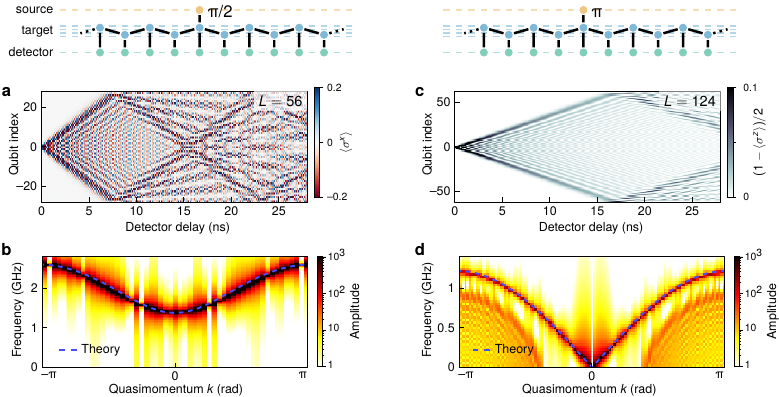}
\caption{\textbf{Single particle approximation predictions.} Simulations are shown using the same parameters as in Fig.~\ref{fig:4} in the main text. {\bf a}, $\sigma^x$ measurement and {\bf c}, excitation-probability outcomes using Eq.~\eqref{eq:m_x_single_photon} and Eq.~\eqref{eq:rho_single_particle} respectively. Their two-dimensional Fourier transforms are shown in {\bf b} and {\bf d} respectively. The predicted peak values of $\omega/(2\pi)$ [given by Eq.~\eqref{eq:omega_sigma_z_single_particle} and Eq.~\eqref{eq:omega_peak_rho_single_particle_1} respectively] are displayed as ``theory'' lines.}
\label{fig:single_particle_pred}
\end{figure*}

For this experiment, we initialized the qubits in the state $\ket{\psi(t=0)} = \ket{0}_e$ and extracted $m_z(n,t)=\bra{\psi(t)} \sigma_n^z \ket{\psi(t)}$ by measuring each qubit in the $\sigma^z$ basis (see Figs.~4c and 4d).
In our approximation, the state at time $t$ is given by $\ket{\psi(t)} = \frac{1}{\sqrt{L}} \sum_k e^{-iE_\text{eff}^{(1)}(k)t/\hbar} \ket{k}_e$.
Restricting to the single excitation sector, we may further write $\sigma_n^z = \sum_m(1-2\delta_{n,m}) \ket{m}_e \bra{m}_e$. Using these expressions, we obtain $m_z(n,t) = 1 - 2 \abs{\eta(n,t)}^2$, where
\begin{align}
\eta(n,t) = 
\frac{1}{L} \sum_{q} e^{i [qn - E_\text{eff}^{(1)}(q)t/\hbar]}.
\end{align}
The excitation probability $\rho$, which equals $(1-m_z)/2$, is therefore given by
\begin{align}
    \rho(n,t) = \abs{\eta(n,t)}^2. \label{eq:rho_single_particle}
\end{align}
Taking the two-dimensional Fourier transform, we get
\begin{align}
    \widetilde \rho(k,\omega) = \frac{2\pi}{L} \sum_{q} \delta(\omega - E_\text{diff}^{k}(q)/\hbar), \label{eq:omega_peak_rho_single_particle}
\end{align}
where $E_\text{diff}^{k}(q) \coloneqq E_\text{eff}^{(1)}(q)-E_\text{eff}^{(1)}(k+q)$.
The peak of the $\widetilde \rho(k,\omega)$ at each $k$ lies at $\omega=E_\text{diff}^{k} (q_*(k))/\hbar$, where $q_*(k)$ denotes the argmin/argmax of $E_\text{diff}^{k}(q)$.
Noting that $q_*(k)=(\pm \pi-k)/2$ using Eq.~\eqref{eq:E_eff}, we conclude that
\begin{align}
\omega_\text{peak}(k) = \pm 2 \mathcal J \sin(\frac{k}{2})/\hbar. \label{eq:omega_peak_rho_single_particle_1}
\end{align}
We indeed find a peak at this value in the experimental data (see Fig.~\ref{fig:4}d in the main text). The predictions (see Fig.~\ref{fig:single_particle_pred}) using the methods of this section compare favorably with the QPU results presented in Fig.~\ref{fig:4} of the main text.

\section{Matrix Product State Simulations}
\label{sec:mps}

We specifically use the yastn library to simulate the evolution of the target system with ideal preparation and detection \cite{YASTN}. The state is evolved according to the time-independent nearest-neighbor target-chain Hamiltonian obtained from Eq.~\eqref{eq:H_anneal} by fixing the anneal parameters to their experimental values, matched to experimental parameters. To match the experiments of Fig.~\ref{fig:4}, the initial MPS is prepared via standard methods (DMRG) as the ground state of the same Hamiltonian with the addition of a polarizing field on the first site $-P \tau^*_0$ (Pauli z or x for $\pi/2$ or $\pi$-pulse preparation respectively), at large P. In effect, site 0 is prepared in a polarized state, and the state of the remaining qubits is a ground state of the length $L-1$ open chain with $h_1=h_{L-1}=\mathcal{J}\langle\tau^z_i\rangle$. This would be a product state in the weak coupling limit, but we do not make such an assumption --- the initialization is therefore slightly more complex than that of Section \ref{sec:single-particle}.

\begin{figure*}[t!]
\includegraphics[scale=1.3]{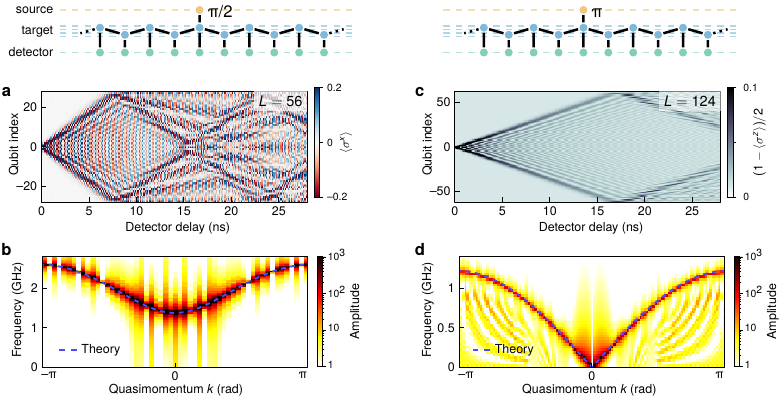}
\caption{\textbf{Matrix product state numerics.} Simulations are shown using the same parameters as in Fig.~4 in the main text. {\bf a}, $\sigma^x$ measurement and {\bf c}, $\sigma^z$ measurements. Their two-dimensional Fourier transforms are shown in {\bf b} and {\bf d} respectively.}
\label{fig:MPS}
\end{figure*}

We then evolve the system by second order 2-site TDVP dynamics and measure single site properties for purposes of Fig.~\ref{fig:MPS}.
Simulation complexity scales inversely with the step size $dt$ (equivalently, linearly with the number of time steps), which controls the second-order Trotterization error, and as $D^3$ with the bond-dimension ($D$), which controls an entanglement-truncation error. At $dt=2^{-6}\SI{}{ns}$ (presented), Trotterization error contributes less to the error than the truncation of bond dimension. We double $D$ until we observe no appreciable changes in the features of the figures presented, which was achieved with $D=16$. For $\pi$ pulse only preparation we can exploit Z2 gauge symmetry for efficiency (or arrive at the same result up to technicalities with the fermionized dynamics of Section \ref{sec:fermionization}).
 
Results are in good agreement with single-excitation approximation presented in the previous section, and with the QPU experiments in spite of experimental simplifications.

A related method was used to verify a subset of the fermionization simulations (Section \ref{sec:fermionization}). Fermionization methods are not subject to Trotterization and bond-dimension errors but are inefficient for general-angle preparations and measurements since modeling is inefficient given symmetry breaking for finite longitudinal fields $h_i$.

The time-dependent variational principle (TDVP) can be applied to matrix product state (MPS) representations of our system to allow scalable simulation of some dynamics~\cite{Cirac-MPSreview}. We apply the methods to demonstrate time evolution of prepared states, with ideal preparation and measurement, and no decoherence.

\section{Simulations for Anderson localization with fermionization}
\label{sec:fermionization}
In one dimension and in the absence of a longitudinal field, the Jordan-Wigner transformation for nearest-neighbor transverse-field Ising models allows for the efficient numerical simulation of dynamics~\cite{Mbengquantum2024, Dziamaraga-exact-solution}. We apply the transformation to demonstrate Anderson localization with $\pi$ pulse initialization and measurement in the energy basis, with ideal preparation and detection, and without decoherence. There is no assumption of a weak coupling limit, or constraint on Hilbert-space sectors, but efficient simulation requires an absence of symmetry breaking effects so that only a restricted set of measurements and preparations are possible.

We can model the imbalance within the Heisenberg picture. Using time-dependent fermionic creation and annihilation operators ($c^\dagger$, $c$ --- we omit the time-dependence for brevity in places), we may substitute
\begin{equation}
\begin{split}
\tau^x_i &= 1 - 2 c_i^\dagger c_i,
\\
\qquad \tau_i^z &= -(c_i + c_i^\dagger) \prod_{j<i}(1-2 c_j^\dagger c _j ), \label{eq:tau_to_c}
\end{split}
\end{equation}
into Eq. \eqref{eq:H_anneal}, to rewrite the spin Hamiltonian as a fermionic Hamiltonian
\begin{equation}
 H = -\frac{1}{2}\sum_i A_i(1 - 2 c_i^\dagger c_i) - \frac{1}{2}\sum_{i=0}^{L-2} J_i (c_i^\dagger - c_i) (c_{i+1} + c_{i+1}^\dagger) - \frac{1}{2} J_{0, L-1}(c_0 + c_0^\dagger)(c_{L-1}^\dagger - c_{L-1}) P \;,
 \label{eq_H_}
\end{equation}
where $h_i=0$, and we abbreviate $J_i = \sqrt{B(s_i(t))B(s_{i+1}(t))}J_{i,i+1}$ and $A_i=A(s_i(t))$.

$P=\prod \tau^x_i$ is the parity operator, and can be taken as $1$ since we apply an even number of $\pi$ pulses (and parity$=1$ is conserved by the dynamics). The quadratic Hamiltonian (\ref{eq_H_}) can be rewritten as
\begin{equation}
H = 
\begin{pmatrix}
\bm{c}^{\dagger}, \bm{c}
\end{pmatrix}
H_{\rm BdG}
\begin{pmatrix}
\bm{c} \\ \bm{c}^{\dagger}
\end{pmatrix},
\end{equation}
where $\bm{c}^{\dagger} = (c_0^{\dagger}, c_1^{\dagger}, \dots, c_{L-1}^{\dagger})$, and $H_{\rm BdG}$ is the known Hermitian $2L \times 2L$ Bogoliubov-de Gennes Hamiltonian in the Nambu space \cite{PhysRev.117.648}, which can be diagonalized via the standard Bogoliubov transformation
\begin{equation}
c_i(t) = \sum_{\mu = 0}^{L-1} 
\big(
u_{i \mu}(t) \gamma_\mu  + v^{\ast}_{i\mu}(t)
\gamma_\mu^\dagger 
\big),
\end{equation}
where $\gamma^{\dagger}_\mu$  ($\gamma_\mu$) creates (annihilates) a Bogoliubov quasiparticle in mode $\mu$, and unitary matrix
\begin{equation}
U = 
\begin{pmatrix}
u & -v^{\ast}
\\
v & u^{\ast}
\end{pmatrix},
\end{equation}
diagonalizes $H_{\rm BdG}$ according to $U^{\dagger} H_{\rm BdG} U = D$. The Heisenberg dynamics can then be simplified
\begin{equation}
[H, c_i] = -A_i c _i + 
\frac{1}{2} \big(
J_i c_{i+1} -J_{i-1} c_{i-1}^\dagger +J_i c_{i+1}^\dagger + J_{i-1} c_{i-1} 
\big),
\label{eq:HBdG}
\end{equation}
up to sign exceptions for boundary spanning couplers.

A product-state initial condition with $\pi$ pulse excitations can be specified by diagonal matrices $u_{ii}(t=0) = 1-v_{ii}(t=0) = 1$ for excited ($\pi$ pulse prepared) states, and $u_{ii}(t=0) = 1 - v_{ii}(t=0) = 0$ for other (ground) states in the weak coupling limit. By substitution into Eq.~\eqref{eq:tau_to_c} and operating on the vacuum state, it can be verified that this yields the correct preparation statistics. For the Anderson localization experiment, the assignment is alternating. Evolving under the experimentally relevant $H_{\rm BdG}$ (without a weak-coupling assumption), we determine the state after time $t$ as
\begin{equation}
\begin{pmatrix}
u(t) \\ v(t)   
\end{pmatrix}
= 
e^{- \mathrm{i} H_{\rm BdG} t/\hbar} 
\begin{pmatrix}
u(0) \\ v(0)   
\end{pmatrix}.
\end{equation}
The desired measurement $\langle \tau^x_i \rangle$ is then achieved by substitution into Eq.~\eqref{eq:tau_to_c}. In Fig.~\ref{fig:5}, we average the expected statistics over 400 independent disorder realizations of the site-dependent fields $A_i$ (equivalently, the detunings $\delta\Delta_i$), with the couplings $J_{ij}$ set according to the compensated experimental parameters.

\end{document}